\documentclass[mlmain]{jmlr}

\usepackage{longtable}

\usepackage[load-configurations=version-1]{siunitx} 

\usepackage{hyperref}       
\usepackage{url}            
\usepackage{amsfonts}       
\usepackage{nicefrac}       
\usepackage{microtype}      
\usepackage{xcolor}         
\usepackage{mathtools}
\usepackage{tikz-cd}

\DeclareMathOperator*{\argmin}{arg\,min}

\DeclareMathOperator*{\bij}{Bij}
\DeclareMathOperator*{\supp}{supp}

\DeclareMathOperator*{\ib}{\text{IB}}
\DeclareMathOperator*{\cl}{\text{cl}}
\newcommand{\iname}{\text{DIB}}
\newcommand{\iib}{\text{IIB}}
\DeclareMathOperator*{\inameop}{\text{\normalfont DIB}}
\newcommand{\ifullname}{Divergence Information Bottleneck }
\newcommand{\Acal}{{\mathcal{A}}}
\newcommand{\Bcal}{{\mathcal{B}}}
\newcommand{\Ecal}{{\mathcal{E}}}
\newcommand{\Lcal}{{\mathcal{L}}}
\newcommand{\Qcal}{{\mathcal{Q}}}
\newcommand{\Scal}{{\mathcal{S}}}
\newcommand{\Tcal}{{\mathcal{T}}}
\newcommand{\Ucal}{{\mathcal{U}}}
\newcommand{\Xcal}{{\mathcal{X}}}
\newcommand{\Ycal}{{\mathcal{Y}}}
\newcommand{\Abar}{{\bar{A}}}

\newcommand{\Xbar}{{\bar{X}}}
\newcommand{\ptilde}{{\tilde{p}}}
\newcommand{\qtilde}{{\tilde{q}}}
\newcommand{\qbar}{{\bar{q}}}

\newcommand{\Acalbar}{{\bar{\mathcal{A}}}}
\newcommand{\Scalbar}{{\bar{\mathcal{S}}}}
\newcommand{\Scaltilde}{{\tilde{\mathcal{S}}}}
\newcommand{\Xcalbar}{{\bar{\mathcal{X}}}}

\newcommand{\piX}{{\pi_{\Xcal}}}
\newcommand{\Gci}{{G_{\text{ci}}}}
\newcommand{\pici}{{\pi_{\text{ci}}}}
\newcommand{\ci}{{\text{ci}}}
\newcommand{\Gce}{{G_{\text{ce}}}}
\newcommand{\pice}{{\pi_{\text{ce}}}}
\newcommand{\ce}{{\text{ce}}}
\newcommand{\Gdi}{{G_{\text{di}}}}
\newcommand{\pidi}{{\pi_{\text{di}}}}
\newcommand{\di}{{\text{di}}}
\newcommand{\Ccong}{C_\text{cong}}
\newcommand{\Dom}{{\text{Dom}}}

\newcommand{\tb}[1]{{#1}}

\jmlrvolume{}
\firstpageno{1}
\editors{Simone Azeglio, Christian Shewmake, Bahareh Tolooshams, Sophia Sanborn, Chase van de Geijin, Nina Miolane}

\jmlryear{2024}
\jmlrworkshop{Symmetry and Geometry in Neural Representations}

\title[Information Parsimony Perspective on Probabilistic Symmetries]{An Informational Parsimony Perspective on \titlebreak Probabilistic Symmetries}

\author{\Name{Hippolyte Charvin} \Email{h.charvin@herts.ac.uk}\\
\Name{Nicola {Catenacci Volpi}} \Email{n.catenacci-volpi@herts.ac.uk}\\
\Name{Daniel Polani} \Email{d.polani@herts.ac.uk}\\
\addr Adaptive Systems Research Group, University of Hertfordshire}

\begin{document}

\maketitle

\begin{abstract}
  Extraction of structure, in particular of group symmetries, is increasingly crucial to understanding and building intelligent models. In particular, some information-theoretic models of parsimonious learning have been argued to induce invariance extraction. Here, we formalise these arguments from a group-theoretic perspective. We then extend them to the study of more general probabilistic symmetries, through compressions preserving geometric measures of complexity. More precisely, \tb{our framework implements} a trade-off between compression and preservation of the divergence from a given hierarchical model, yielding a novel generalisation of the Information Bottleneck framework. Through appropriate choices of hierarchical models, we fully characterise (in the discrete and full support case) channel invariance, channel equivariance and distribution invariance under permutation. Allowing imperfect divergence preservation then leads to principled definitions of ``soft symmetries'', where the ``coarseness'' corresponds to the degree of compression of the system. In simple synthetic experiments, we demonstrate that our method successively recovers, at increasingly compressed ``resolutions'', nested but increasingly perturbed equivariances, where new equivariances emerge at bifurcation points of the trade-off parameter. Our framework suggests a new path for the extraction of generalised probabilistic symmetries.
\end{abstract}

\section{Introduction}

Group symmetries have become highly relevant to the study of intelligence, from the structure of sensory processing \citep{kellerSpacetimePerspectiveDynamical2024} and its relation to an agent's actions \citep{godonFormalAccountStructuring2020,keurtiStitchingManifoldsLeveraging2024} to equivariant neural networks \citep{gerkenGeometricDeepLearning2023} and structure-discovering AI models \citep{vanderouderaaNoethersRazorLearning2024}. This relevance is often understood as a consequence of the pervasiveness of symmetries in the natural world: biological and artificial systems interacting with such a highly structured environment should \emph{leverage} its symmetries, e.g., by internalising them into their own information-processing. But what leverage do symmetries provide exactly? Intuitively, the presence of symmetries in a system allows for a \emph{simpler} description of it --- thus potentially improving the efficiency of learning and generalisation about this system. In other words, a system's symmetries afford the possibility of \emph{informationally parsimonious} descriptions of it. 

More precisely, the \emph{projection on orbits} of a symmetry group's action can be seen as an information-preserving compression, as it preserves the information about anything invariant under the group action. This suggests that projections on orbits might be solutions to well-chosen rate-distortion problems, hence opening the way to the integration of group symmetries into an information-theoretic framework. \tb{If successful, such an} integration could formalise the link between symmetry and information parsimony, but also $(i)$ yield natural ways to ``soften'' group symmetries into flexible concepts more relevant to real-world data \tb{--- which often lacks exact symmetries despite exhibiting a strong ``structure'' --- and $(ii)$ enable} symmetry discovery through the optimisation of information-theoretic quantities.

As a first step in this direction, we introduce a novel rate-distortion-inspired framework \tb{whose solutions} capture, or at least mimick, the projection on orbits of certain group symmetries. We call it Divergence Information Bottleneck (\iname), as it generalises the Information Bottleneck (IB) method \citep{tishbyInformationBottleneckMethod2000}.
Here, the encoders trade-off compressing data with preserving its divergence from a given hierarchical model, potentially under constraints on \tb{the} encoder shape. As divergences from hierarchical models are geometric measures of complexity \citep{ayGeometricApproachComplexity2011}, this setting produces \emph{complexity-preserving} compressions. With \tb{appropriate} choices of hierarchical models and shape constraints, we obtain encoders which, for full divergence preservation, characterise various group-theoretic symmetries. Allowing only \emph{partial} divergence preservation then leads to principled definitions of \emph{soft symmetries}, whose \tb{``softness'' is parametrised by the compression-divergence preservation trade-off: i.e., softer symmetries correspond to more compression and less preserved divergence.}

The IB method has previously been argued to extract invariances \citep{achilleEmergenceInvarianceDisentanglement2018a}.
In Section \ref{section:ib_invariance}, we prove that it indeed characterises group-theoretic channel invariances. This motivates the DIB framework, which we specialise to characterise the equivariances of a channel $p(Y|X)$ and the invariances of a distribution $p(A)$ (Section \ref{section:sib}). 
We then present simple synthetic numerical experiments on soft equivariances (Section \ref{section:numerical_experiments}), \tb{where we} study channels satisfying a series of nested equivariances that have been perturbed to various degrees. We show that our framework recovers the perturbed equivariances, at successive bifurcation points of the trade-off parameter corresponding to increasingly compressed resolutions. Finally, we present the limitations of our approach (Section \ref{section:limitations}), and conclude in Section \ref{section:conclusion}.




\paragraph{Assumptions and notations}

All alphabets are finite, except bottleneck alphabets $\Tcal := \mathbb{N}$. The probability simplex defined by an alphabet $\Acal$ is denoted by $\Delta_{\Acal}$. The set of conditional probabilities, also called channels, from $\mathcal{A}$ to $\mathcal{B}$, resp. to $\mathcal{A}$ itself, is denoted by $C(\mathcal{A}, \mathcal{B})$, resp. $C(\mathcal{A})$. Functions are seen as deterministic channels. Channels are often seen as linear maps between vector spaces of measures, where the image of the distribution $p$ through the channel $\gamma$ is written $\gamma \cdot p$. By extension, $f \cdot a := f(a)$ for an element $a$ and a function $f$.
The symbol $\circ$ denotes channel composition. The set of bijections of $\mathcal{A}$ is $\bij(\mathcal{A})$, the uniform distribution $\Ucal(\Acal)$, the identity map  $e_{\mathcal{A}}$, and $\Scal^c := \Acal \setminus \Scal$ for $\Scal \subseteq \Acal$. For $p_1 \in \Delta_\Acal, p_2 \in \Delta_\Bcal$, their \emph{tensor product} is defined through $(p_1 \otimes p_2)(a,b) := p_1(a) p_2(b)$. Similarly $(\mu \otimes \eta)(b,b'|a,a') := \mu(b|a) \eta(b'|a')$ for $\mu \in C(\Acal, \Bcal)$, $\eta \in C(\Acal', \Bcal')$. 

\section{Information Bottleneck and Group Invariances}
\label{section:ib_invariance}

The IB implements a trade-off between compressing a variable, say $X$, and preserving the information that $X$ carries about a second variable $Y$. More precisely, let $p(X,Y) \in \Delta_{\Xcal \times \Ycal}$ with $p(X)$ full-support. The corresponding IB problem is \citep{gilad-bachrachInformationTheoreticTradeoff2003}
\begin{align} \label{eq:ib_problem_primal}
    \ib(\lambda) := \argmin_{\substack{q(T|X) \, \in \, C(\mathcal{X}, \mathcal{T}) \; : \\ I_q(T;Y) \geq \lambda}} \; I_q(X;T),
\end{align}
where $q(X,Y,T) := p(X,Y)q(T|X)$ and $0 \leq \lambda \leq \Lambda := I(X;Y)$.

On the other hand, the \emph{channel invariance group} $G_{\text{ci}}$ of $p(Y|X)$ is the group of $\sigma \in \bij(\Xcal)$ such that $p(Y|X) \circ \sigma = p(Y|X)$, with projection on orbits written $\pici : \Xcal \rightarrow \Xcal/G_\ci$. Crucially, here $\pici$ characterises $\Gci$: it can be easily verified that $\sigma \in \Gci \Leftrightarrow \pici \circ \sigma = \pici$.

We now want to show that the solutions $\kappa \in \ib(\Lambda)$ essentially coincide with $\pici$, thus yielding an information-theoretic characterisation of $\Gci$ through such $\kappa$. We will need the equivalence relation
\begin{align} \label{eq:def:equivalence_relation_ib}
  x \sim_{\Xcal} x' \quad \Leftrightarrow \quad p(Y|x) = p(Y|x'),
\end{align}
with corresponding partition $\Xcalbar$ and projection $\pi_\Xcal : \Xcal \rightarrow \Xcalbar$; along with the following notion.



\begin{definition}    \label{def:congruent_channel}
  The set of \emph{congruent channels} \citep{ayInformationGeometry2017} from $\Acal$ to $\Bcal$, denoted by $\Ccong(\Acal, \Bcal)$, is that of channels $\gamma$ such that there exists a function $f : \Bcal \rightarrow \Acal$ with $f \circ \gamma = e_\Acal$.
\end{definition}

In particular, composing an encoder $\kappa$ with a congruent channel $\gamma$ can be seen as a trivial operation, in that the output of $\kappa$ can be unambigously recovered from that of $\gamma \circ \kappa$.

\begin{theorem} \label{th:results_ib}
  For $\Lambda := I(X;Y)$ and all $\sigma \in \bij(\Xcal)$, the following holds:
  \begin{enumerate}
    \item $\ib(\Lambda) = \left\{ \gamma \circ \pi_{\Xcal} \, : \ \ \gamma \in \Ccong(\Xbar, \Tcal) \right\}$.
    \item Let $\kappa \in \ib(\Lambda)$. Then  $\sigma \in \Gci$ if and only if $\kappa \circ \sigma = \kappa$.
    \item If $\sigma \in \Gci$, then $\kappa \circ \sigma = \kappa$ also holds for all $0 \leq \lambda \leq \Lambda$ and $\kappa \in \ib(\lambda)$.
    \item The projection $\pi_{\Xcal}$ defined by $\sim_{\Xcal}$  coincides with the projection on orbits $\pici$.
  \end{enumerate}
\end{theorem}
\begin{proof}
  See Appendices \ref{section:apx:general_stuff} and \ref{section:apx:proof_results_ib}.
\end{proof}

Crucially, point $(ii)$ means that invariances are thoses bijections $\sigma$ such that the effect of transforming $\Xcal$ with $\sigma$ is ``quotiented out'' by $\kappa \in \ib(\Lambda)$. Point $(i)$ explains why: these $\kappa$ implement precisely (up to trivial transformations) the quotient of $\Xcal$ by the equivalence relation $\sim_\Xcal$ that equates elements of $\Xcal$ providing the same information about $Y$ (see \eqref{eq:def:equivalence_relation_ib}). 
Point $(iii)$ shows that the ``quotienting out'' of invariances by bottlenecks also occurs for all values of the trade-off parameter $\lambda$, even though it is only a full characterisation for $\lambda = I(X;Y)$.
Point $(iv)$, combined with point $(i)$, means that the projection $\pici$, defined purely in group-theoretic terms, is characterised as the solution to the zero-distortion case of a generalised rate-distortion problem, here the IB \citep{zaidiInformationBottleneckProblems2020}. 
Note that point $(i)$ is redundant with existing results \citep{shamirLearningGeneralizationInformation2010};\footnote{Point $(i)$ can be seen as the fact that $\ib(\Lambda)$ consists of minimal sufficient statistics of $X$ w.r.t. $Y$, proven in \citep{shamirLearningGeneralizationInformation2010}. But our new proof also yields point $(iii)$ and mirrors that of Theorem~\ref{th:results_sib} below.} and point $(ii)$ is not surprising as previous work already linked the IB method to invariance extraction \citep{achilleEmergenceInvarianceDisentanglement2018a}. However, our group-theoretic formalisation provides guidance for generalisations of this phenomenon: i.e., for reformulating and softening probabilistic symmetries with the language of information theory. The following sections provide first steps in this direction. 

\section{Divergence Information Bottleneck and Group Symmetries}
\label{section:sib}

\subsection{General framework}
\label{section:sib_general}

Fix a distribution $p = p(A) \in \Delta_\Acal$, an exponential family $\Ecal \subseteq \Delta_\Acal$, and a subset of encoders $C \subseteq C(\Acal, \Tcal)$. We then define the \emph{\ifullname} (DIB) as
\begin{align}   \label{eq:def:sib}
    \iname(\lambda) := \argmin_{\substack{\kappa \in C \\ D(\kappa \cdot p || \kappa \cdot \Ecal) \geq \lambda}} \, I_\kappa(A;T),
\end{align}
where $0 \leq \lambda \leq \Lambda := D(p||\Ecal)$, and, denoting by $\cl \Ecal$ the topological closure of $\Ecal$ in $\Delta_\Acal$,
\begin{align}
  D(p||\Ecal) &:= \inf_{r \in \cl \Ecal} \, D\big(p(A)||r(A)\big) = D\big(p(A)||\ptilde(A)\big), \label{eq:def_divergence_p_from_exp_family} \\
  D( \kappa \cdot p || \kappa \cdot \Ecal) &:= \inf_{r \in \cl \Ecal} \, D\big((\kappa \cdot p)(T)||(\kappa \cdot r)(T)\big) = D\big((\kappa \cdot p)(T) || (\kappa \cdot \ptilde)(T)\big). \label{eq:def_divergence_p_from_exp_family_latent_space}
\end{align}
Here $\ptilde \in \cl \Ecal$ is the unique distribution which \tb{simultaneously} achieves the minimum in \eqref{eq:def_divergence_p_from_exp_family} and \eqref{eq:def_divergence_p_from_exp_family_latent_space} (see Appendix \ref{section:apx:sib_on_projection_exponential}). While $D(p||\Ecal)$ is the divergence of $p$ from $\Ecal$, here $D( \kappa \cdot p || \kappa \cdot \Ecal)$ measures the ``divergence of $p$ from $\Ecal$ \emph{in the latent space} $\Tcal$, through the lens of the channel $\kappa$''. Solutions to \eqref{eq:def:sib} can thus be seen as optimal compressions of $A$ under the constraint of (partially or fully) preserving the divergence of $p(A)$ from the exponential family $\Ecal$.
The choice of $C$ allows to potentially enforce constraints on the shape of encoder channels $\kappa$.

Intuitively, $D(p||\Ecal)$ measures the presence of a specific structure in $p(A)$, formalised as the divergence from the family $\Ecal$ of distributions which do not have such structure. E.g., for $\Acal = \Xcal \times \Ycal$ and $\Ecal := \Delta_{\Xcal} \otimes \Delta_{\Ycal}$, we have $D(p(X,Y)||\Ecal) = I(X;Y)$: the corresponding DIB (with e.g. $C = C(\Xcal \times \Ycal,\Tcal)$) is a \tb{\emph{mutual information-preserving}} joint compression of $X$ and $Y$ \citep{charvinInformationTheoryBasedDiscovery2023a}. More generally, the divergence from a \emph{hierarchical model} $\Ecal$ measures the complexity of a system's given set of interdependencies \citep{ayGeometricApproachComplexity2011}. The DIB is tailored for this setting, where solutions to \eqref{eq:def:sib} are thus \emph{complexity-preserving optimal compressions}. However, Theorem~\ref{th:results_sib} below holds for any exponential family $\Ecal$.

Let us define, on $\Acal$, the equivalence relation
\begin{align}   \label{eq:def:equivalence_relation_sib}
    a \sim a' \quad \Leftrightarrow \quad p(a) \ptilde(a') = p(a') \ptilde(a),
\end{align}
with corresponding partition $\Acalbar := \{\Acal_j\}_{j=1,\dots,n}$ and projection $\pi : \Acal \rightarrow \bar{\Acal}$. 

\begin{theorem} \label{th:results_sib}
  If $C = C(\Acal, \Tcal)$ and $\supp(p(A)) = \Acal$, then $\iname(\Lambda) = \left\{ \gamma \circ \pi \, : \ \gamma \in \Ccong(\Abar, \Tcal) \right\}$.
\end{theorem}


I.e., for full support $p(A)$ and no constraints on the shape of encoders, the fully divergence-preserving solutions $\kappa \in \iname(\Lambda)$ coincide, up to trivial transformations, with the clustering of $\Acal$ defined by the relation \eqref{eq:def:equivalence_relation_sib}.
The theorem is proven in Appendices \ref{section:apx:general_stuff} and \ref{section:apx:proofs_sib}.

\subsection{Application to equivariances}
\label{section:sib_applied_to_equivariances}

Consider now $\Acal = \Xcal \times \Ycal$ equipped with a full support distribution $p(X,Y)$; in particular, $p(Y|X)$ is well-defined. The group $\Gce$ of \emph{channel equivariances} is the group of pairs $(\sigma, \tau) \in \bij(\Xcal) \times \bij(\Ycal)$ such that $p(Y|X) \circ \sigma = \tau \circ p(Y|X)$. Crucially, \tb{as was the case} for invariances (see Section \ref{section:ib_invariance}), the projection on orbits $\pice$ characterises the group $\Gce$: it can be easily verified that $(\sigma, \tau) \in \Gce \Leftrightarrow \pice \circ (\sigma \otimes \tau) = \pice$.

We want to design a DIB problem that mimics this property of $\pice$. Note that here $\pice$ does not compress $\Xcal$ and $\Ycal$ separately but \emph{jointly} \citep{charvinInformationTheoryBasedDiscovery2023a}, so that it is natural to impose no constraint on encoders' shape: $C = C(\Xcal \times \Ycal, \Tcal)$. Moreover, it can be verified\footnote{See Lemma 15 in \citep{charvinInformationTheoryBasedDiscovery2023a}.} that $\pice(x,y) = \pice(x',y')$ implies $p(y|x) = p(y'|x')$. Based on this observation, we search for an exponential family $\Ecal$ such that the relation $\sim$ from equation \eqref{eq:def:equivalence_relation_sib} becomes 
\begin{align}  \label{eq:def:equivalence_relation_sib_equivariances}
  (x,y) \sim (x',y') \quad &\Leftrightarrow \quad p(y|x) = p(y'|x').
\end{align}
This is achieved by choosing
\begin{align*}
  \Ecal = \Ecal_\ce :=  \{ r(X)\Ucal(\Ycal), \ r(X) \in \Delta_{\Xcal} \},
\end{align*}
which yields $\ptilde(X,Y) = p(X)\Ucal(\Ycal)$, so that the relation $p(x,y)\ptilde(x',y') = p(x',y') \ptilde(x,y)$ (see equation~\eqref{eq:def:equivalence_relation_sib}) here means $p(y|x) = p(y'|x')$.
Note that $\Ecal_\ce$ coincides with the hierarchical model of probabilities on $\Xcal \times \Ycal$ that actually depend only on $\Xcal$. Borrowing from the geometric approach to complexity \citep{ayGeometricApproachComplexity2011}, we thus interpret $D(p(X,Y)||\Ecal_\ce)$ as the ``degree to which the system $(X,Y)$ is more than just $X$'', or equivalently, ``the amount of information that $p(Y|X)$ carries about $(X,Y)$''.
More precisely, $\Ecal_\ce$ is the family of probabilities $r(X,Y)$ such that $r(Y|x)$ (when well-defined) is always equal to the maximum entropy distribution $ \Ucal(\Ycal)$. This condition means, intuitively, \tb{that the channel $r(Y |X)$ is ``maximally uninformative'', i.e., that it ``carries no information'' about the system $(X, Y)$. Through the decomposition $r(X,Y) = r(X)r(Y|X)$, this is equivalent to the intuition that all
the information about $(X, Y )$ is contained in the marginal $r(X)$.}

The divergence $D(p||\Ecal_\ce)$ then quantifies ``how far'' \tb{$p = p(X,Y)$} is from such distributions. Thus the set $\iname_\ce(\lambda)$ of solutions to the $\iname$ problem \eqref{eq:def:sib} with $\Acal = \Xcal \times \Ycal$, $\Ecal = \Ecal_\ce$ and parameter $\lambda$ is that of optimal compressions \emph{preserving the information carried by the channel $p(Y|X)$} (to the degree $\lambda$).

\begin{theorem} \label{th:charac_equivariances_with_sib}
  The following holds, for $\Lambda := D(p||\Ecal_\ce)$ and all $(\sigma, \tau) \in \bij(\Xcal) \times \bij(\Ycal)$\emph{:}\footnote{Appendix \ref{section:apx:relation_to_iib} clarifies how \tb{this work relates to our previous results} in \citep{charvinInformationTheoryBasedDiscovery2023a}.}
  \begin{enumerate}
    \item Let $\kappa \in \emph{\iname}_\ce(\Lambda)$. Then $(\sigma, \tau) \in \Gce$ if and only if $\kappa \circ (\sigma \otimes \tau) = \kappa$.
    \item If $(\sigma, \tau) \in \Gce$, then $\kappa \circ (\sigma \otimes \tau) = \kappa$ also holds for all $0 \leq \lambda \leq \Lambda$ and $\kappa \in \emph{\iname}_\ce(\lambda)$.
    \item The projection $\pi$ defined by $\sim$ in equation \eqref{eq:def:equivalence_relation_sib_equivariances} does not, in general, coincide with $\pice$.
  \end{enumerate}
\end{theorem}

See Appendix \ref{section:apx:proof_charac_equivariances_with_sib} for the proof, which relies on Theorem~\ref{th:results_sib}.
Most importantly, point $(i)$ means that equivariances of $p(Y|X)$ are those pairs of transformations $(\sigma, \tau)$ such that the effect of simultaneously transforming $\Xcal$ with $\sigma$ and $\Ycal$ with $\tau$ is ``quotiented out'' by the coarse-grainings $\kappa \in \text{DIB} (\Lambda)$, making these transformations indiscernible from the identity. 
The $\iname_\ce$ framework thus provide an \emph{information-theoretic characterisation of
equivariances}. \tb{It} can be summarised \tb{with the following equivalence of commutative diagrams:}
\begin{align*}
  \begin{tikzcd}[ampersand replacement=\&, column sep=large, row sep = large]
      \Xcal \arrow[d, swap,"\sigma"] \arrow[r, "p(Y|X)"] \& \Ycal \arrow[d,"\tau"]\\
      \Xcal \arrow[r,"p(Y|X)"]      \&  \Ycal
  \end{tikzcd}
      \quad \quad \Leftrightarrow \quad \quad 
  \begin{tikzcd}[ampersand replacement=\&, column sep=tiny, row sep = large]
      \Xcal \times \Ycal\arrow[rr,"\sigma \otimes \tau"] \arrow[dr, swap, "\kappa"] \&  \& \Xcal \times \Ycal \arrow[dl,"\kappa"]\\
          \& \Tcal \& 
  \end{tikzcd}
\end{align*}
\tb{which means that the commutativity of the left-hand-side  diagram --- i.e., the defining property of equivariances --- is equivalent to that of the right-hand-side one --- an information-theoretic property.} Moreover, point $(ii)$ in Theorem~\ref{th:charac_equivariances_with_sib} means that the ``quotienting out'' of equivariances happens actually for all granularities $\lambda$, even though \tb{the equivalence only holds} for $\lambda = \Lambda$. 
Point $(iii)$ says that, unfortunately, the clustering $\pi$ obtained from $\iname_\ce(\Lambda)$ does not always coincide with the projection on orbits $\pice$ \tb{(see Appendix \ref{section:apx:proof_charac_equivariances_with_sib} for more details).}

\smallskip

We can now draw upon our new information parsimony perspective on equivariances to soften this group-theoretic notion, where each granularity $\lambda$ defines a corresponding set of soft equivariances. Let $0 \leq \lambda \leq \Lambda$. We define a \emph{$\lambda$-equivariance} as a pair $(\mu, \eta) \in C(\Xcal) \otimes C(\Ycal)$ such that there exists $\kappa \in \iname_\ce(\lambda)$ with $\kappa \circ (\mu \otimes \eta) = \kappa$.  In other words, a soft equivariance is defined through the very same equation $\kappa \circ (\sigma \otimes \tau) = \kappa$ that characterises exact equivariances, but where the fully information-preserving compression $\kappa$ is now only a \emph{partially} information-preserving compression. Moreover, we allow $\mu$ and $\eta$ to be non-invertible and stochastic.

\smallskip

To conclude this section, let us point out that the classic IB can be recovered as a DIB with the same exponential family $\Ecal_\ce$ as for equivariances, but with shape constraints $C \subsetneq C(\Acal,\Tcal)$ which impose that $\kappa$ can only compress $\Xcal$ and not $\Ycal$. See Appendix \ref{section:apx:ib_is_sib}.



\subsection{Application to distribution invariances}
\label{section:sib_applied_to_distrib_invariances}

We now proceed similarly for transformations leaving a given distribution invariant. I.e., let $p \in \Delta_\Acal$ be full support, and define the group $\Gdi$ of \emph{distribution invariances} as the group of $\Phi \in \bij(\Acal)$ such that $p(\Phi \cdot A) = p(A)$. As we do not consider any structure on $\Acal$, we choose \tb{unconstrained encoders}, i.e., $C = C(\Acal, \Tcal)$. Moreover, as $\Phi \in \Gdi$ if and only if $p(a) = p(\Phi \cdot a)$ for all $a$, it is natural to search for an exponential family yielding the equivalence relation $a \sim a' \Leftrightarrow p(a) = p(a')$. It can be easily verified that this is achieved by choosing \tb{only} the uniform distribution: $\Ecal = \Ecal_\di := \{\Ucal(\Acal)\}$. Intuitively, here the $\iname$ problem, which we denote by $\iname_\di$, preserves (partially or wholly) the divergence $D(p(A)||\Ucal(\Acal))$ of $p(A)$ from the uniform distribution: i.e., it preserves the ``degree to which $p(A)$ is deterministic".

\begin{theorem} \label{th:charac_distrib_invariances_with_sib}
  The following holds, for $\Lambda := D(p||\Ecal_\di)$ and all $\Phi \in \bij(\Acal)$:
  \begin{enumerate}
    \item Let $\kappa \in \emph{\iname}_\di(\Lambda)$. Then $\Phi \in \Gdi$ if and only if $\kappa \circ \Phi = \kappa$.
    \item If $\Phi \in \Gci$, then $\kappa \circ \Phi = \kappa$ also holds for all $0 \leq \lambda \leq \Lambda$ and $\kappa \in \emph{\iname}_\di(\lambda)$.
    \item The projection $\pi$ defined by $\sim$ coincides with the projection on orbits $\pidi$.
  \end{enumerate}
\end{theorem}


The proof relies on Theorem~\ref{th:results_sib} (see Appendix \ref{section:apx:proof_charac_distrib_invariances_with_sib}). Interpretations of points $(i)$ and $(ii)$ are analogous to those for equivariances.
Point $(iii)$ highlights that here, $\pidi$ and $\pi$ do coincide. Eventually, one can directly adapt the definition of soft equivariances to one for soft distribution invariances.

\subsection{Relevant computational and conceptual tools}
\label{section:sib_tools}

Here, we present an algorithm \tb{approximating solutions to the DIB \eqref{eq:def:sib} for unconstrained encoders, and two relevant concepts.} Consider the Lagrangian relaxation of \eqref{eq:def:sib},
\begin{align} \label{eq:def:sib_lagrangian}
  \argmin_{\kappa \in C} \ \Big[ I_\kappa(A;T) - \beta D(\kappa \cdot p || \kappa \cdot \ptilde) \Big],
\end{align} 
where $\beta \geq 0$. For $C = C(\Acal, \Tcal)$,  deriving the Lagrangian above w.r.t. $\kappa$ yields a fixed-point equation that all local minimisers must satisfy (similarly to the classic IB). From this fixed-point equation, we obtain a Blahut-Arimoto-like (BA) algorithm with the same guarantees as BA for the classic IB \citep{tishbyInformationBottleneckMethod2000} (see Appendices \ref{section:apx:blahut_arimoto_extend_from_support} and \ref{section:apx:blahut_arimoto}).\footnote{The convergence speed (and thus computational feasability) of this and other BA algorithms is difficult to estimate --- particularly as they exhibit critical slowing down near bifurcations \citep{agmonCriticalSlowingTopological2021a}.} In the following, we will write $\kappa_\beta$ for the output of the BA algorithm (i.e., a local minimser) with parameter $\beta$, and also $I_\beta := I_{\kappa_\beta}(A;T)$ and $D_\beta := D(\kappa_\beta \cdot p || \kappa_\beta \cdot \ptilde)$. Note that both $I_\beta$ and $D_\beta$ increase with $\beta$.

Now the \emph{effective cardinality} \citep{zaslavskyDeterministicAnnealingEvolution2019} of some  $\kappa \in \iname(\lambda)$ is defined as the minimum number of symbols $t$ necessary to describe the output of $\kappa$ (see Appendix \ref{section:apx:charac_effcard_with_frac} for a formal definition). In all our numerical experiments, we observed that: $(i)$ similarly as for the classic IB, effective cardinality monotically increases with $\beta$, and $(ii)$ changes of effective cardinality coincide with discontinuities in the slope of the curve $\beta \mapsto (I_\beta, D_\beta)$, which is reminiscent of the second-order bifurcations observed for the IB \citep{zaslavskyDeterministicAnnealingEvolution2019}. We will thus here refer to changes of effective cardinalities as bifurcations.

Eventually, we want to investigate whether the equation $\kappa_\beta \circ \Phi = \kappa_\beta$ is satisfied for varying $\beta$ and varying $\Phi \in G$, with $G$ some fixed subgroup of $\bij(\Acal)$. But numerically, it is also important, when this equation is not exactly satisfied, to quantify the extent of the deviation. We propose to use the divergence defined for all channel $\kappa \in C(\Acal, \Tcal)$ as
\begin{align*}
  D^p(\kappa || C_G) :=  \min_{\nu \in C_G} \,  D^p(\kappa || \nu) :=  \min_{\nu \in C_G} \, \sum_{a \in \supp(p(A))} p(a) D(\kappa(T|a) || \nu(T|a)),
\end{align*}
where $C_G := \{ \nu : \, \forall \Phi \in G, \, \nu \circ \Phi = \nu \}$ is the family of channels that are exactly input-symmetric w.r.t $G$. Intuitively, $D^p(\kappa || C_G)$ measures the divergence of the channel $\kappa$ from being input-symmetric for the action of $G$ on the distribution $p(A)$. In particular, $D^p(\kappa || C_G) = 0$ if and only if $\kappa \circ \Phi = \kappa$ for all $\Phi \in G$.  See Appendix \ref{section:apx:computation_channel_divergence} for more details.

\section{Numerical experiments}
\label{section:numerical_experiments}

\subsection{Synthetic numerical experiments on equivariances}
\label{section:numerical_experiments_equivariance}

\begin{figure}
  \centering
  \includegraphics[scale=0.32]{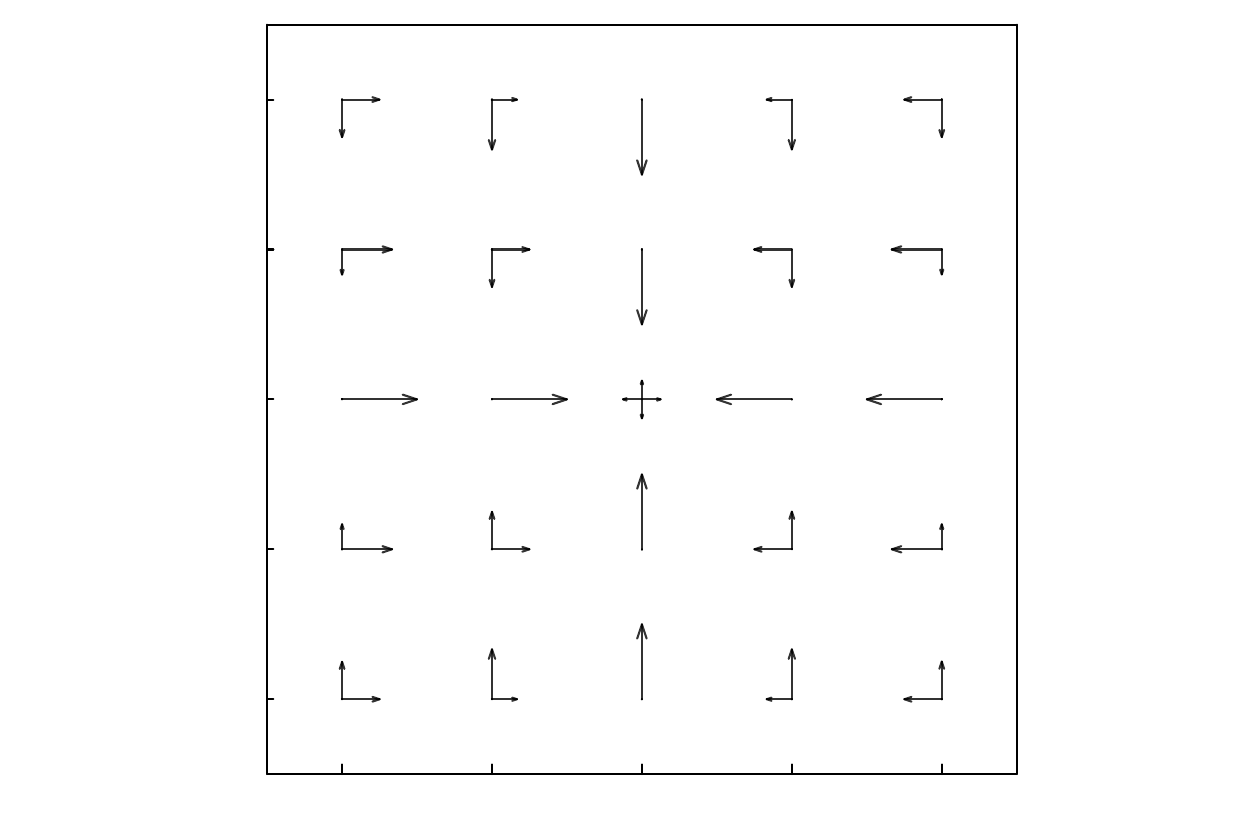}
  \includegraphics[scale=0.32]{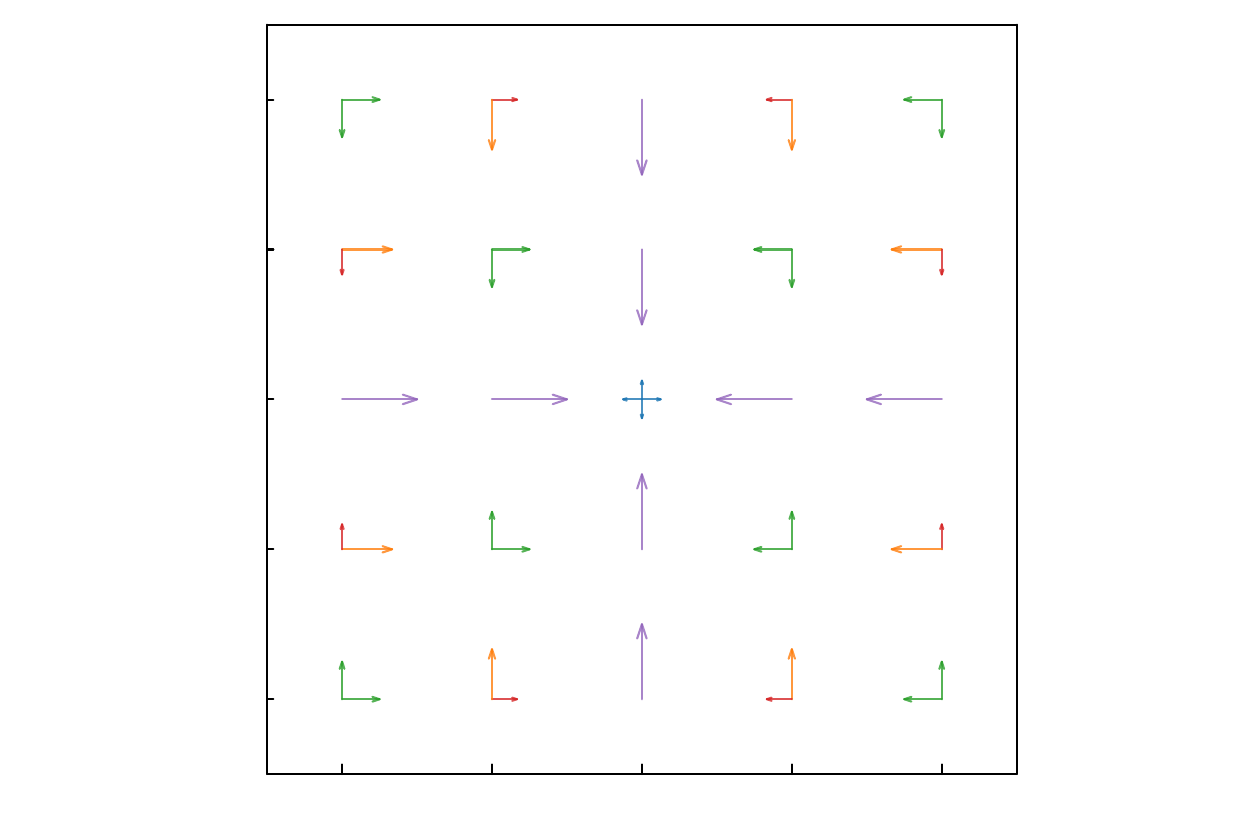}
  \caption{Left: representation of $p(Y|X)$, where $X$ is the position on the grid, $Y$ the gradient direction, and probabilities are proportional to arrow lengths. \tb{Thus equivariances are here pairs $(\sigma, \tau)$ that send each arrow on an arrow of equal length.}
  Right: same figure with colors representing a clustering of $\supp(p(X,Y))$ --- which defines a clustering of $\Xcal \times \Ycal$ if we add the cluster $\supp(p(X,Y))^c$, made of position-orientation pairs with probability $0$, i.e., with no arrow. The latter clustering is obtained in 2 distinct ways: $(i)$ as the projection on orbits of the equivariance group of $p(Y|X)$, and $(ii)$ as the clustering defined by relation \eqref{eq:def:equivalence_relation_sib_equivariances}.}
  \label{fig:proba_grid}
\end{figure}

The concept of soft $\lambda$-equivariance from Section \ref{section:sib_applied_to_equivariances} is motivated by the case of full information preservation $\lambda = \Lambda$, where \tb{our new definition of $\lambda$-equivariances coincides} with that of classic group equivariances. However, it is not a priori clear that, once $\lambda < \Lambda$, our generalisation \tb{is consistent with the intuition of approximate symmetry}. Here, we provide a sanity check in this direction, in a simple synthetic grid-world scenario. We start from an exactly equivariant channel $p(Y|X)$, which we synthetically perturb in such a way to render some of its equivariances more perturbed than others. \tb{We then expect that $(i)$ all the perturbed equivariances should be recovered as soft $\lambda$-equivariances by the $\iname_\ce$ framework, once the reduction of the parameter $\lambda$ enforces sufficient compression, and $(ii)$ more perturbed equivariances should need more compression before being recovered as soft equivariances --- because it means that more noise needs to be ``overlooked''}. 

We compute the $\iname_\ce$ solutions with the BA algorithm described in Appendix \ref{section:apx:blahut_arimoto}, combined with reverse deterministic annealing, starting from $T = (X,Y)$ for large $\beta$ (similarly as for the classic IB in \citep{zaslavskyDeterministicAnnealingEvolution2019}). All our experiments were simulated on a PC having 32GB of RAM and a 2.3GHz 12th generation i7 CPU.

Here, $\Xcal$ stands for positions on a $5 \times 5$ grid, and $\Ycal$ for a gradient with 4 possible directions. Thus $p(Y|X)$ describes the probability of a direction at a given position, which can be thought of, e.g., as a nutrient gradient sensed by a bacteria. We choose uniform $p(X)$ (choosing non-uniform $p(X)$ resulted in similar results). As seen in Figure \ref{fig:proba_grid}, left, $p(Y|X)$ has many symmetries: it can be verified that the equivariance group of $p(Y|X)$ has 6 distinct orbits (one is $\supp(p(X,Y))^c$), represented in Figure \ref{fig:proba_grid}, right. Moreover, even though we saw in Theorem \ref{th:charac_equivariances_with_sib}, point $(iii)$, that the projection on orbits $\pice$ and the clustering $\pi$ defined by relation \eqref{eq:def:equivalence_relation_sib_equivariances} do not generally coincide, here they do coincide. Thus Figure \ref{fig:proba_grid}, right, also represents $\pi$.

\begin{figure}
  \centering
  \includegraphics[scale=0.32]{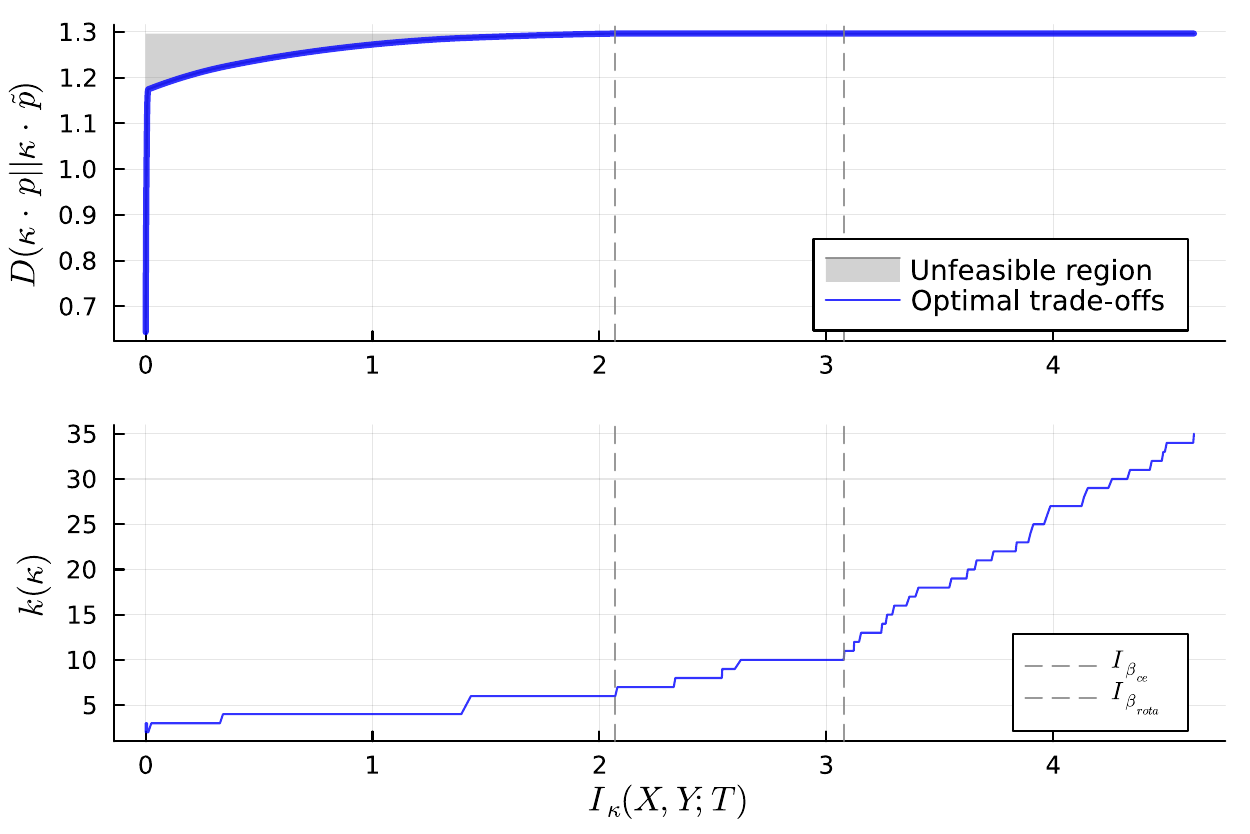}
  \includegraphics[scale=0.32]{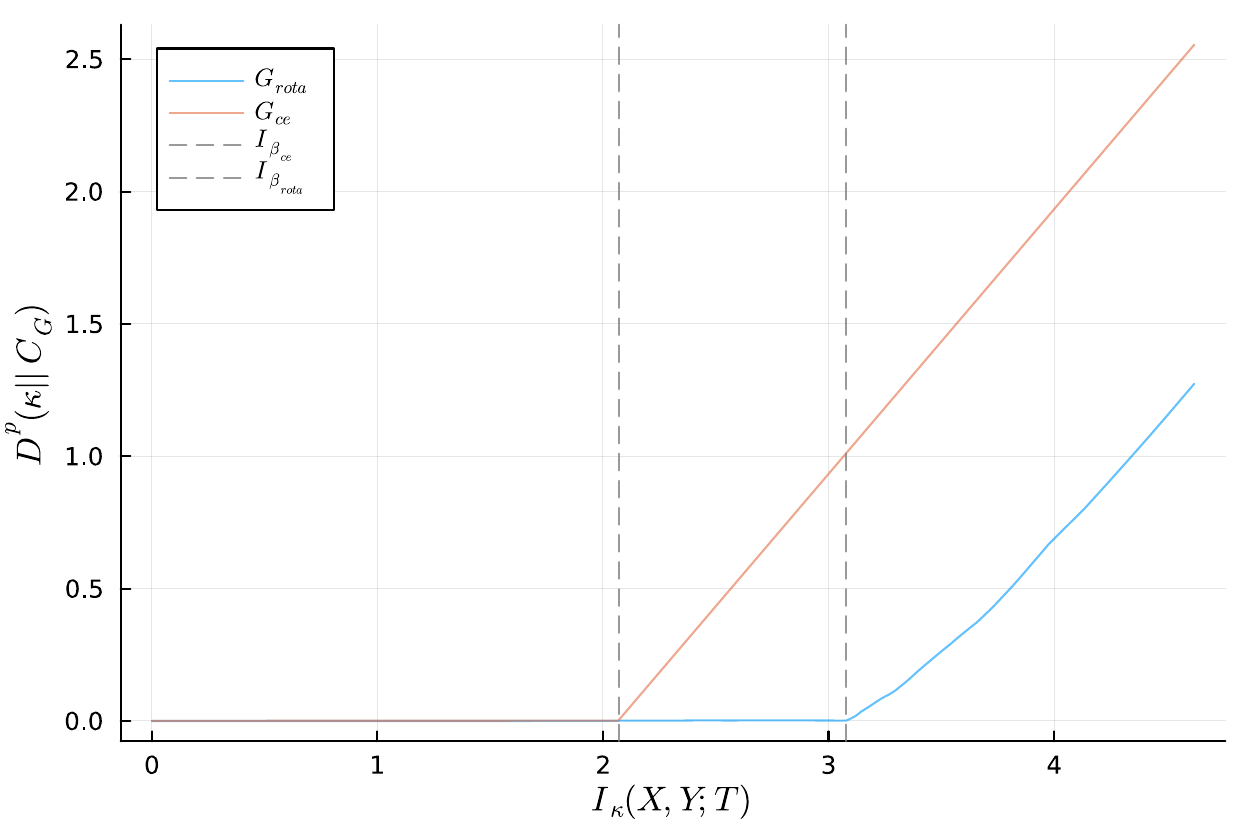}
  \caption{$D_\beta := D(\kappa_\beta \cdot p || \kappa_\beta \cdot \ptilde)$ as a function of $I_\beta := I_{\kappa_\beta}(X,Y;T)$.  Bottom left: Effective cardinality $k(\kappa)$ as a function of $I_\beta$. Right: Divergence of compression channels $\kappa_\beta$ as a function of $I_\beta$, for the groups $G_{\ce}$ and $G_{\text{rota}}$. The vertical dashed lines represent specific bifurcations of the parameter $\beta$ at which $D^p(\kappa_\beta||C_{G_{\text{rota}}})$, resp. $D^p(\kappa_\beta||C_{G_\ce})$, approximately vanishes (in decreasing order of $I_\beta$).}
    \label{fig:information}
\end{figure}

From Section \ref{section:sib_tools}, we have $D^p(\kappa_\beta||C_{G_\ce}) = 0$ if and only if $\kappa_\beta \circ (\sigma, \tau) = \kappa_\beta$ for all $(\sigma, \tau) \in \Gce$.
Theorem \ref{th:charac_equivariances_with_sib}, point $(ii)$, suggests that this equation \tb{may} indeed hold for all $\beta$.\footnote{Here, the full support assumption, which is required in Theorem \ref{th:charac_equivariances_with_sib}, does not hold for $p(X,Y)$. We leave to future work the theoretical study of the case $\supp(p(X,Y)) \subsetneq \Xcal \times \Ycal$.} As a sanity check, we thus computed the $\iname_\ce$ bottlenecks $\kappa_\beta$ for $0 \leq D_\beta \leq \Lambda$, and indeed obtained $D^p(\kappa_\beta||C_{G_\ce}) \leq 3 \cdot 10^{-16}$ for all $\beta$. We also noted that the bottlenecks' effective cardinality monotonically increases from 1 for $D_\beta = 0$ to 6 for $D_\beta = \Lambda$.

We then perturb $p(Y|X)$ with two random perturbations. The first one, of larger amplitude, breaks some equivariances in $G_\ce$, but not all of them. More precisely, after the perturbation, $p(Y|X)$ still satisfies the equivariances from its subgroup $G_{\text{rota}} \subsetneq G_\ce$ generated by rotating both the positions and the gradient directions by $90$ degrees. The second perturbation applied to $p(Y|X)$, of smaller amplitude, breaks all the remaining equivariances from $G_{\text{rota}}$. We thus obtain a new $p(Y|X)$ which, intuitively, is still ``approximately'' equivariant, but where the approximate equivariances in $\Gce \setminus G_{\text{rota}}$ are \emph{coarser} than those in $G_{\text{rota}}$, because the perturbation was larger for the former than for the latter.

We compute 1000 $\iname_\ce$-bottlenecks for varying $\beta$ (which took 339 seconds).
The resulting information curve $(I_\beta,D_\beta)$, along with the corresponding effective cardinalities, are shown in Figure \ref{fig:information}, left. As for the classic IB, we obtain a non-decreasing and concave information curve, and an increasing effective cardinality (except for small $I_\beta$, which could be due to numerical errors).

Crucially, we then observe (Figure \ref{fig:information}, right) that for decreasing $\beta$, the divergences $D^p(\kappa_\beta||C_{G_{\text{rota}}})$ and $D^p(\kappa_\beta||C_{G_\ce})$ successively vanish, at bifurcation values $\beta_{\text{rota}}$, resp. $\beta_\ce < \beta_{\text{rota}}$. Thus the perturbed equivariances are here recovered  by the $\iname_\ce$ method as \emph{soft} equivariances, for large enough compression. Moreover, as the equivariances from $G_{\text{rota}}$ have been less perturbed that those in the remaining of $G_{\ce}$, here the degree of compression required to recover an approximate equivariance scales with the ``coarseness'' of that equivariance.

Eventually, note that, in Figure \ref{fig:information}, left, the gain in divergence $D_\beta$ from $I_\beta = I_{\beta_\ce}$ to the maximum value $I_{\text{max}}$ of $I_\beta$ is negligible, whereas $I_{\text{max}} - I_{\beta_\ce}$ is large. This resonates with the intuition that underlying symmetries in raw data afford a potentially drastic informational compression ($I_\beta$ here), under a negligible loss in informational accuracy ($D_\beta$ here). 





\section{Limitations}
\label{section:limitations}

Our core results are of theoretical nature, and hold in the discrete and full support case. At this stage, it is still unclear whether and how they extend to continuous and non fully supported distributions. Numerically, 
the BA class of algorithms addresses only the discrete case and generally scales unfavourably in larger scenarios. Future work could make the DIB problem amenable to deep network optimisation by adapting the classic IB's variational bounds \citep{alemiDeepVariationalInformation2017}. 

Moreover, the DIB corresponding to equivariances does not yield the projection on  orbits under the group's action (Theorem~\ref{th:charac_equivariances_with_sib}, \tb{point $(iii)$}). Our $\iname_\ce$ framework still characterises group equivariances for maximum parameter $\lambda = \Lambda$, but for $\lambda < \Lambda$, this limitation might make the concept of $\lambda$-equivariance ill-adapted to the kind of ``soft equivariances'' that we hope to identify in real-world data. Future work should test further the scientific relevance of $\lambda$-equivariances as defined here, and potentially design a new instance or variation of the DIB which would capture the projection on orbits of the equivariance group. The link between our soft equivariances and other proposals of generalised equivariances \citep{wangApproximatelyEquivariantNetworks2022a,romeroLearningPartialEquivariances2022, songFlowFactorizedRepresentation2023,ashmanApproximatelyEquivariantNeural2024} should also be clarified.

Let us stress that at the current stage, our method provides an information-preserving compression $\kappa$ corresponding to --- or mimicking --- the symmetry group's projection on orbits, but does not discover the symmetries themselves. However, our ``projection on orbits''-oriented perspective grounds group symmetries in IB theory, which could serve as a building block for information theory-based symmetry discovery.


\section{Conclusion}
\label{section:conclusion}

Motivated by the ability of the classic IB to implicitly extract channel invariances, we investigated generalizations of this phenomenon. For this, we introduce the Divergence IB, a novel and substantial generalization of the classic IB. We show how this method can generalize the informational characterisation of invariances to that of channel equivariances and distribution invariances. Crucially, expressing these symmetries through IB-like trade-offs yields a natural softening of these very stringent group-theoretic notions. This suggests a principled route to extract the data's underlying symmetries through \emph{complexity-preserving coarse-grainings} of it, thus exposing the data's ``platonic core'', so to say.

However, while we only investigated some canonical examples of symmetries, the DIB framework is highly versatile. With other exponetial families $\Ecal$ and channel shape constraints $C$, this method could help discover novel kinds of scientifically relevant structures.

Eventually, our work suggests a new path to formalise the intuition that informationally parsimonious systems, e.g., embodied agents, should leverage the coarse symmetries underlying the interaction with their environment.


\bibliography{my_library.bib}

\begin{thebibliography}{23}
\providecommand{\natexlab}[1]{#1}
\providecommand{\url}[1]{\texttt{#1}}
\expandafter\ifx\csname urlstyle\endcsname\relax
  \providecommand{\doi}[1]{doi: #1}\else
  \providecommand{\doi}{doi: \begingroup \urlstyle{rm}\Url}\fi

\bibitem[Achille and Soatto(2018)]{achilleEmergenceInvarianceDisentanglement2018a}
Alessandro Achille and Stefano Soatto.
\newblock Emergence of {{Invariance}} and {{Disentanglement}} in {{Deep Representations}}.
\newblock In \emph{2018 {{Information Theory}} and {{Applications Workshop}} ({{ITA}})}, pages 1--9, February 2018.
\newblock \doi{10.1109/ITA.2018.8503149}.

\bibitem[Agmon et~al.(2021)Agmon, Benger, Ordentlich, and Tishby]{agmonCriticalSlowingTopological2021a}
Shlomi Agmon, Etam Benger, Or~Ordentlich, and Naftali Tishby.
\newblock Critical {{Slowing Down Near Topological Transitions}} in {{Rate-Distortion Problems}}.
\newblock In \emph{2021 {{IEEE International Symposium}} on {{Information Theory}} ({{ISIT}})}, pages 2625--2630, July 2021.
\newblock \doi{10.1109/ISIT45174.2021.9517956}.

\bibitem[Alemi et~al.(2017)Alemi, Fischer, Dillon, and Murphy]{alemiDeepVariationalInformation2017}
Alexander~A. Alemi, Ian Fischer, Joshua~V. Dillon, and Kevin Murphy.
\newblock Deep {{Variational Information Bottleneck}}.
\newblock In \emph{International {{Conference}} on {{Learning Representations}}}, February 2017.

\bibitem[Ashman et~al.(2024)Ashman, Diaconu, Weller, Bruinsma, and Turner]{ashmanApproximatelyEquivariantNeural2024}
Matthew Ashman, Cristiana Diaconu, Adrian Weller, Wessel Bruinsma, and Richard~E. Turner.
\newblock Approximately {{Equivariant Neural Processes}}.
\newblock \emph{Advances in Neural Information Processing Systems}, 37:\penalty0 97088--97123, December 2024.

\bibitem[Ay(2015)]{ayInformationGeometryComplexity2015}
Nihat Ay.
\newblock Information {{Geometry}} on {{Complexity}} and {{Stochastic Interaction}}.
\newblock \emph{Entropy}, 17\penalty0 (4):\penalty0 2432--2458, April 2015.
\newblock ISSN 1099-4300.
\newblock \doi{10.3390/e17042432}.

\bibitem[Ay et~al.(2011)Ay, Olbrich, Bertschinger, and Jost]{ayGeometricApproachComplexity2011}
Nihat Ay, Eckehard Olbrich, Nils Bertschinger, and J{\"u}rgen Jost.
\newblock A geometric approach to complexity.
\newblock \emph{Chaos: An Interdisciplinary Journal of Nonlinear Science}, 21\penalty0 (3):\penalty0 037103, September 2011.
\newblock ISSN 1054-1500.
\newblock \doi{10.1063/1.3638446}.

\bibitem[Ay et~al.(2017)Ay, Jost, L{\^e}, and Schwachh{\"o}fer]{ayInformationGeometry2017}
Nihat Ay, J{\"u}rgen Jost, H{\^o}ng~V{\^a}n L{\^e}, and Lorenz Schwachh{\"o}fer.
\newblock \emph{Information {{Geometry}}}, volume~64 of \emph{Ergebnisse Der {{Mathematik}} Und Ihrer {{Grenzgebiete}} 34}.
\newblock Springer International Publishing, Cham, 2017.
\newblock ISBN 978-3-319-56477-7 978-3-319-56478-4.
\newblock \doi{10.1007/978-3-319-56478-4}.

\bibitem[Charvin et~al.(2023)Charvin, Volpi, and Polani]{charvinInformationTheoryBasedDiscovery2023a}
Hippolyte Charvin, Nicola~Catenacci Volpi, and Daniel Polani.
\newblock Towards {{Information Theory-Based Discovery}} of {{Equivariances}}.
\newblock In \emph{{{NeurIPS}} 2023 {{Workshop}} on {{Symmetry}} and {{Geometry}} in {{Neural Representations}}}, November 2023.

\bibitem[Csisz{\'a}r and K{\"o}rner(2011)]{csiszarInformationTheoryCoding2011}
Imre Csisz{\'a}r and J{\'a}nos K{\"o}rner.
\newblock \emph{Information {{Theory}}: {{Coding Theorems}} for {{Discrete Memoryless Systems}}}.
\newblock Cambridge University Press, Cambridge, 2 edition, 2011.
\newblock ISBN 978-0-521-19681-9.
\newblock \doi{10.1017/CBO9780511921889}.

\bibitem[Gerken et~al.(2023)Gerken, Aronsson, Carlsson, Linander, Ohlsson, Petersson, and Persson]{gerkenGeometricDeepLearning2023}
Jan~E. Gerken, Jimmy Aronsson, Oscar Carlsson, Hampus Linander, Fredrik Ohlsson, Christoffer Petersson, and Daniel Persson.
\newblock Geometric deep learning and equivariant neural networks.
\newblock \emph{Artificial Intelligence Review}, 56\penalty0 (12):\penalty0 14605--14662, December 2023.
\newblock ISSN 1573-7462.
\newblock \doi{10.1007/s10462-023-10502-7}.

\bibitem[{Gilad-Bachrach} et~al.(2003){Gilad-Bachrach}, Navot, and Tishby]{gilad-bachrachInformationTheoreticTradeoff2003}
Ran {Gilad-Bachrach}, Amir Navot, and Naftali Tishby.
\newblock An {{Information Theoretic Tradeoff}} between {{Complexity}} and {{Accuracy}}.
\newblock In Gerhard Goos, Juris Hartmanis, Jan Van~Leeuwen, Bernhard Sch{\"o}lkopf, and Manfred~K. Warmuth, editors, \emph{Learning {{Theory}} and {{Kernel Machines}}}, volume 2777, pages 595--609. Springer Berlin Heidelberg, Berlin, Heidelberg, 2003.
\newblock ISBN 978-3-540-40720-1 978-3-540-45167-9.
\newblock \doi{10.1007/978-3-540-45167-9_43}.

\bibitem[Godon et~al.(2020)Godon, Argentieri, and Gas]{godonFormalAccountStructuring2020}
Jean-Merwan Godon, Sylvain Argentieri, and Bruno Gas.
\newblock A {{Formal Account}} of {{Structuring Motor Actions With Sensory Prediction}} for a {{Naive Agent}}.
\newblock \emph{Frontiers in Robotics and AI}, 7, 2020.
\newblock ISSN 2296-9144.
\newblock \doi{10.3389/frobt.2020.561660}.

\bibitem[Keller et~al.(2024)Keller, Muller, Sejnowski, and Welling]{kellerSpacetimePerspectiveDynamical2024}
T.~Anderson Keller, Lyle Muller, Terrence~J. Sejnowski, and Max Welling.
\newblock A {{Spacetime Perspective}} on {{Dynamical Computation}} in {{Neural Information Processing Systems}}, September 2024.
\newblock Preprint at \url{https://arxiv.org/abs/2409.13669}.

\bibitem[Keurti et~al.(2024)Keurti, Sch{\"o}lkopf, Aceituno, and Grewe]{keurtiStitchingManifoldsLeveraging2024}
Hamza Keurti, Bernhard Sch{\"o}lkopf, Pau~Vilimelis Aceituno, and Benjamin~F. Grewe.
\newblock Stitching {{Manifolds}}: {{Leveraging Interaction}} to {{Compose Object Representations}} into {{Scenes}}.
\newblock In \emph{{{ICML}} 2024 {{Workshop}} on {{Geometry-grounded Representation Learning}} and {{Generative Modeling}}}, June 2024.

\bibitem[Romero and Lohit(2022)]{romeroLearningPartialEquivariances2022}
David~W. Romero and Suhas Lohit.
\newblock Learning {{Partial Equivariances From Data}}.
\newblock \emph{Advances in Neural Information Processing Systems}, 35:\penalty0 36466--36478, December 2022.

\bibitem[Shamir et~al.(2010)Shamir, Sabato, and Tishby]{shamirLearningGeneralizationInformation2010}
Ohad Shamir, Sivan Sabato, and Naftali Tishby.
\newblock Learning and generalization with the information bottleneck.
\newblock \emph{Theoretical Computer Science}, 411\penalty0 (29):\penalty0 2696--2711, 2010.
\newblock ISSN 0304-3975.
\newblock \doi{10.1016/j.tcs.2010.04.006}.

\bibitem[Song et~al.(2023)Song, Keller, Sebe, and Welling]{songFlowFactorizedRepresentation2023}
Yue Song, Andy Keller, Nicu Sebe, and Max Welling.
\newblock Flow {{Factorized Representation Learning}}.
\newblock \emph{Advances in Neural Information Processing Systems}, 36:\penalty0 49761--49782, December 2023.

\bibitem[Tishby et~al.(2000)Tishby, Pereira, and Bialek]{tishbyInformationBottleneckMethod2000}
Naftali Tishby, Fernando~C. Pereira, and William Bialek.
\newblock The information bottleneck method, April 2000.
\newblock Preprint at \url{https://arxiv.org/abs/physics/0004057}.

\bibitem[{van der Ouderaa} et~al.(2024){van der Ouderaa}, {van der Wilk}, and {de Haan}]{vanderouderaaNoethersRazorLearning2024}
Tycho~F. {van der Ouderaa}, Mark {van der Wilk}, and Pim {de Haan}.
\newblock Noether's {{Razor}}: {{Learning Conserved Quantities}}.
\newblock \emph{Advances in Neural Information Processing Systems}, 37:\penalty0 135943--135965, December 2024.

\bibitem[Wang et~al.(2022)Wang, Walters, and Yu]{wangApproximatelyEquivariantNetworks2022a}
Rui Wang, Robin Walters, and Rose Yu.
\newblock Approximately {{Equivariant Networks}} for {{Imperfectly Symmetric Dynamics}}.
\newblock In \emph{Proceedings of the 39th {{International Conference}} on {{Machine Learning}}}, pages 23078--23091. PMLR, June 2022.

\bibitem[Yeung(2008)]{yeungInformationTheoryNetwork2008}
Raymond~W. Yeung.
\newblock \emph{Information {{Theory}} and {{Network Coding}}}.
\newblock Springer, 2008.

\bibitem[Zaidi et~al.(2020)Zaidi, {Estella-Aguerri}, and Shamai~(Shitz)]{zaidiInformationBottleneckProblems2020}
Abdellatif Zaidi, I{\~n}aki {Estella-Aguerri}, and Shlomo Shamai~(Shitz).
\newblock On the {{Information Bottleneck Problems}}: {{Models}}, {{Connections}}, {{Applications}} and {{Information Theoretic Views}}.
\newblock \emph{Entropy}, 22\penalty0 (2), 2020.
\newblock ISSN 1099-4300.
\newblock \doi{10.3390/e22020151}.

\bibitem[Zaslavsky and Tishby(2019)]{zaslavskyDeterministicAnnealingEvolution2019}
Noga Zaslavsky and Naftali Tishby.
\newblock Deterministic annealing and the evolution of {{Information Bottleneck}} representations.
\newblock August 2019.
\newblock Preprint at \url{https://www.nogsky.com/publication/2019-evo-ib/2019-evo-IB.pdf}.

\end{thebibliography}

\appendix

\section{General notations, definitions and results}
\label{section:apx:general_stuff}






The proofs of Theorem \ref{th:results_ib} (see Appendix \ref{section:apx:proof_results_ib}) and Theorem \ref{th:results_sib} (see Appendix \ref{section:apx:proofs_sib}.) are very similar. Here, we collect definitions and pieces of reasoning that are common to both. 

We fix a finite set $\Acal$, a probability $p \in \Delta_{\Acal}$ which we assume full support,
and a channel $\kappa = q(T|A) \in C(\Acal, \Tcal)$. We also consider a partition $\Acalbar = \{\Acal_j\}_{j=1,\dots,n}$ of $\Acal$, and denote by $\pi : \Acal \rightarrow \Acalbar$ the corresponding projection. Whenever it can simplify notations, we will identify $\Acalbar$ with $\{1, \dots, n\}$. We associate to $\kappa = q(T|A)$ a corresponding channel $\bar{\kappa} = \qbar(T|\Acal_J)\in C(\Acalbar, \Tcal)$ defined through
\begin{align} \label{eq:def:qbar}
  \qbar(t|j) := \qbar(t|\Acal_j) := q(t|\Acal_j) := \frac{\sum_{a \in \Acal_j} q(t|a) p(a)}{p(\Acal_j)}.
\end{align}
Intuitively, $\qbar(T|\Acal_J)$ is the ``channel induced by $q(T|A)$ (and $p(A)$) when quotienting its input space $\Acal$ into the partition $\Acalbar$''. We also define a channel $\kappa_\pi = q_\pi(T|A) \in C(\Acal, \Tcal)$ through,
for all $a \in \Acal$, $t \in \Tcal$,
\begin{align} \label{eq:def:q_pi}
  q_\pi(t|a) := ( \bar{\kappa} \circ \pi ) (t|a) = \sum_j q(t|\Acal_j) \delta_{a \in \Acal_j}.
\end{align}
Intuitively, $q_\pi(T|A)$ is the ``enforced factorisation of $q(T|A)$ through $\pi$''. Indeed: $(i)$ it is a channel defined from $q(T|A)$ that factorises through $\pi$; and $(ii)$, whenever $q(T|A)$ itself factorises through $\pi$, then we must have $q(T|A) = q_\pi(T|A)$ (see point $(i)$ in Lemma \ref{lemma:general:gamma_is_gammaq_and_I_is_HpiA}). 

\begin{lemma} \label{lemma:factorisation_decreases_info}
  We have
  \begin{enumerate}
    \item $q_\pi(T) = q(T)$.
    \item $I_{q_\pi}(A;T) \leq I_q(A;T)$, where equality holds if and only if $q_\pi(T|A) = q(T|A)$. 
  \end{enumerate}
\end{lemma}

\begin{proof}
  $(i)$. For all $t \in \Tcal$,
  \begin{align*}
    q_\pi(t) &= \sum_{a \in \Acal} p(a) q_\pi(t|a)\\
    &= \sum_j \sum_{a \in \Acal_j} p(a) q_\pi(t|a) \\
    &= \sum_j \sum_{a \in \Acal_j} p(a) q(t|\Acal_j) \\
    &= \sum_j p(\Acal_j) q(t|\Acal_j) \\
    &= \sum_j \sum_{a \in \Acal_j} q(t|a) p(a) = q(t),
  \end{align*}
  where the second and last equalities use the fact that $\{ \Acal_j \}_j$ is a partition of $\Acal$, and the fourth uses the definition of $q(t|\Acal_j)$ in \eqref{eq:def:qbar}.

  $(ii)$. We have
  \begin{align*}
      I_q(A;T) &= \sum_{a \in \Acal, t \in \supp(q(T))} p(a) q(t|a) \log \left( \frac{q(t|a)}{q(t)} \right) \\
      &= \sum_{t \in \supp(q(T))} \sum_j \sum_{a \in \Acal_j} p(a) q(t|a) \log \left( \frac{q(t|a)}{q(t)} \right).
  \end{align*}
  But from the log-sum inequality \citep{csiszarInformationTheoryCoding2011}, for fixed $t \in \supp(q(T))$ and fixed $j$,
  \begin{align} \label{eq:local:logsum_ineq_I_q_qprime}
      \sum_{a \in \Acal_j} p(a) q(t|a) \log \left( \frac{q(t|a)}{q(t)} \right) &= \sum_{a \in \Acal_j} p(a) q(t|a) \log \left( \frac{p(a)q(t|a)}{p(a)q(t)} \right)\notag\\
      &\geq \left( \sum_{a \in \Acal_j} p(a) q(t|a) \right) \log \left( \frac{\sum_{a \in \Acal_j} p(a) q(t|a)}{\sum_{a \in \Acal_j} p(a)q(t)} \right) \\
      &= p(\Acal_j) q(t|\Acal_j) \log \left( \frac{q(t|\Acal_j)}{q(t)} \right),  \notag
  \end{align}
  with equality in \eqref{eq:local:logsum_ineq_I_q_qprime} if and only if $\frac{q(t|a)p(a)}{q(t)p(a)}$ is constant for $a \in \Acal_j$, i.e., if and only if $q(t|a)$ is constant for $a \in \Acal_j$. Note that the last line of \eqref{eq:local:logsum_ineq_I_q_qprime} can be rewritten
  \begin{align*}
    p(\Acal_j) q(t|\Acal_j) \log \left( \frac{q(t|\Acal_j)}{q(t)} \right)  &= \sum_{a \in \Acal_j} p(a) q_\pi(t|a) \log \left( \frac{q_\pi(t|a)}{q(t)} \right) \\
    &= \sum_{a \in \Acal_j} p(a) q_\pi(t|a) \log \left( \frac{q_\pi(t|a)}{q_\pi(t)} \right),
  \end{align*}
  where the second equality uses point $(i)$ proven above. Thus, summing \eqref{eq:local:logsum_ineq_I_q_qprime} over $j$ and $t \in \supp(q(T))$, we get
  \begin{align*}
    I_q(A;T) &\geq  \sum_{t \in \supp(q(T))} \sum_j \sum_{a \in \Acal_j} p(a) q_\pi(t|a) \log \left( \frac{q_\pi(t|a)}{q_\pi(t)} \right)  \\
    &= I_{q_\pi}(A;T),
  \end{align*}
  with equality if and only if for all $t \in \supp(q(T))$ and all $j$, the quantity $q(t|a)$ is constant for $a \in \Acal_j$ --- which means more precisely that for all $a \in \Acal_j$, we have $q(t|a) = q(t|\Acal_j)$. In other words, there is equality if and only if $q(t|a) = q_\pi(t|a)$ for all $a \in \Acal$ and $t \in \supp(q(T))$. As the definition of $q_\pi(T|A)$ clearly implies $\supp(q(T)) = \supp(q_\pi(T))$, the latter is equivalent to $q(T|A) = q_\pi(T|A)$.
\end{proof}

We now introduce, for all $t \in \Tcal$, the set
\begin{align} \label{eq:def:Scaltq}
  \Acal_t^q := \left\{ a \in \Acal : \ \ q(t|a) > 0 \right\},
\end{align}
which can be seen as the ``probabilistic pre-image of $t$ through the channel $q(T|A)$''.

\begin{lemma} \label{lemma:general:charac_congruent}
  The following are equivalent:
  \begin{enumerate}
    \item The channel $\qbar(T|\Acal_J) \in C(\Acalbar, \Tcal)$ defined in \eqref{eq:def:qbar} is congruent.
    \item For all $t \in \supp(q(T))$, there exists a partition element $\Acal_j \in \Acalbar$ such that  $\Acal_t^q \subseteq \Acal_j$.
  \end{enumerate}
\end{lemma}
Note that the $\Acal_j$ satisfying  $\Acal_t^q \subseteq \Acal_j$ is actually unique, because $\{\Acal_j\}_j$ is a partition of $\Acal$, and $\Acal_t^q \neq \emptyset$ for $t \in \supp(q(T))$.

\begin{proof}
  For any function $f : \Tcal \rightarrow \{1,\dots,n\}$ and all $j,j'$,
  \begin{align*}
    (f \circ \qbar)(j'|j) &= \sum_{t \in \Tcal} \delta_{j'=f(t)} \qbar(t|j) = \sum_{t \in \Tcal} \delta_{j'=f(t)} q(t|\Acal_j) = q(f^{-1}(j')|\Acal_j),
  \end{align*}
  where $f^{-1}(j')$ is the pre-image of $j'$ through $f$. Thus
  \begin{align}
    (f \circ \qbar)(j'|j) = \delta_{j'=j} \quad &\Leftrightarrow \quad q(f^{-1}(j')|\Acal_j) = \delta_{j'=j} \\
    &\Leftrightarrow \quad q(f^{-1}(j')|\Acal_j) > 0 \text{ only if } j' = j  \label{eq:ib:loc:constraint_iff_congruent_1} \\
    &\Leftrightarrow \quad  \forall t \in \supp(q(T)), \ \ q(t|\Acal_j) > 0 \text{ only if } f(t) = j \label{eq:ib:loc:constraint_iff_congruent_2} \\
    &\Leftrightarrow \quad \forall t \in \supp(q(T)), \ \ \left( \exists a \in \Acal_j \, : \, q(t|a) > 0 \right) \text{ only if } f(t) = j \label{eq:ib:loc:constraint_iff_congruent_3}\\
    &\Leftrightarrow \quad \forall t \in \supp(q(T)), \ \ \left( \Acal_t^q \cap \Acal_j \neq \emptyset \right) \text{ only if } f(t) = j \label{eq:ib:loc:constraint_iff_congruent_4}
  \end{align}
  where line \eqref{eq:ib:loc:constraint_iff_congruent_1} uses the fact that $q(\cdot|\Acal_j)$ is a probability measure and $\{f^{-1}(j)\}_j$ a partition of $\Tcal$; line \eqref{eq:ib:loc:constraint_iff_congruent_2} that $q(f^{-1}(j')|\Acal_j) = \sum_{t : f(t) = j'} q(t|\Acal_j)$; line \eqref{eq:ib:loc:constraint_iff_congruent_3} that $q(t|\Acal_j) = \frac{1}{p(\Acal_j)} \sum_{a \in \Acal_j}q(t|a)p(a)$ with $\Acal_j \subseteq \Acal$; and line \eqref{eq:ib:loc:constraint_iff_congruent_4} the definition of $\Acal_t^q$. Therefore,
  \begin{align}
    \exists f : \Tcal & \rightarrow \{1,\dots,n\}, \ \forall j,j', \ (f \circ \qbar)(j'|j) = \delta_{j'=j} \label{eq:ib:loc:constraint_iff_congruent_5}  \\
    & \Leftrightarrow \quad
    \exists f : \Tcal \rightarrow \{1,\dots,n\}, \ \forall j, \forall t \in \supp(q(T)), \  \left( \Acal_t^q \cap \Acal_j \neq \emptyset \right) \text{ only if } f(t) = j \notag \\
    &\Leftrightarrow \quad \exists f : \Tcal \rightarrow \{1,\dots,n\}, \ \forall t \in \supp(q(T)), \ \Acal_t^q \subseteq \Acal_{f(t)} \label{eq:ib:loc:constraint_iff_congruent_6}
  \end{align}
  But on the one hand, the statement \eqref{eq:ib:loc:constraint_iff_congruent_5} is the definition of $\qbar(T|\Acal_J)$ being a congruent channel. On the other hand, because any function on $\supp(q(T))$ can be arbitrarily extended to the whole $\Tcal$, the statement \eqref{eq:ib:loc:constraint_iff_congruent_6} is equivalent to 
  \begin{align*}
    \exists f : \supp(q(T)) \rightarrow \{1,\dots,n\}, \ \forall t \in \supp(q(T)), \ \Acal_t^q \subseteq \Acal_{f(t)},
  \end{align*}
  which is clearly a reformulation of point $(ii)$.
\end{proof}

\begin{lemma} \label{lemma:general:gamma_is_gammaq_and_I_is_HpiA}
  Fix a channel $\gamma \in C(\Acalbar, \Tcal)$, and define $\kappa := q(T|A) := \gamma \circ {\pi}$. Then
  \begin{enumerate}
    \item $\gamma$ coincides with the channel $\bar{\kappa} := \qbar(T|\Acal_J)$ defined in equation \eqref{eq:def:qbar}.
    \item If moreover $\gamma$ is congruent, then $I_{q}(A;T) = H({\pi}(A))$.
  \end{enumerate}
\end{lemma}
\begin{proof}
  $(i)$. For all $\Acal_j \in \Acalbar$,
  \begin{align*}
    \qbar(t|\Acal_j) := \sum_{a \in \Acal_j} \frac{p(a)q(t|a)}{p(\Acal_j)} = \sum_{a \in \Acal_j} \frac{p(a)\gamma(t|{\pi}(a))}{p(\Acal_j)} = \sum_{a \in \Acal_j} \frac{p(a)\gamma(t|\Acal_j)}{p(\Acal_j)} = \gamma(t|\Acal_j).
  \end{align*}

  $(ii)$. Fix a deterministic function $f : \Tcal \rightarrow \Acalbar$ such that $f \circ \gamma = e_{\Acalbar}$. Then 
  \begin{align*}
    I_q(A;T) \geq I_q(A;f(T)) = I_q(A; f \circ \gamma (\pi(A))) = I(A;\pi(A)) = H(\pi(A)),
  \end{align*}
  where the last equality holds because $\pi$ is deterministic. On the other hand, as a direct consequence of the factorisation $q(T|A) = \gamma \circ \pi$, we have the Markov chain $A - {\pi}(A) - T$. Thus $I_q(A;T) \leq I(A;\pi(A)) = H(\pi(A))$.
\end{proof}

\section{Proof of Theorem \ref{th:results_ib}}
\label{section:apx:proof_results_ib}

The following sections present the successive steps of the proof. Let us first note that all the definitions and statements from Appendix \ref{section:apx:general_stuff} can be applied here, with $\Acal = \Xcal$ and $p(A) = p(X)$ (we assumed that $p(A)$ is full support in Appendix \ref{section:apx:general_stuff}, but also that $p(X)$ is full support in Section \ref{section:ib_invariance}). 
We write here $\Acalbar = \Xcalbar = \{\Xcal_j\}_{j=1,\dots,n}$ and $\pi = \pi_{\Xcal} : \Xcal \rightarrow \Xcalbar$ (these notations coincide with those defined in Section \ref{section:ib_invariance}); also $q(T|A) = q(T|X)$, $\qbar(T|\Acal_J) = \qbar(T|\Xcal_J)$, $q_\pi(T|A) = q_{\pi_{\Xcal}}(T|X)$. Moreover, recall that here $q(T|X)$ defines not only a joint distribution $q(X,T)$, but also, together with $p(X,Y)$, a joint distribution $q(X,Y,T)$ through the assumed Markov chain $T - X - Y$. Similarly, $q_\pi(T|X)$ defines a joint distribution $q_\pi(X,Y,T)$.

\subsection{Factorisation for all parameter $\lambda$}
\label{section:apx:proof_ib_factorises_for_all_lambda}

\begin{proposition} \label{prop:ib_factorises_for_all_lambda}
  For all $0 \leq \lambda \leq \Lambda$, every solution $q(T|X) \in \ib(\lambda)$ factorises as $q(T|X) = \qbar(T|\Xcal_J) \circ {\piX}$.
\end{proposition}

The proof will derive from the following lemma:

\begin{lemma} \label{lemma:ib:relation_q_qprime}
  For every $q(T|X)$, we have:
  \begin{enumerate}
    \item $q_{\piX}(Y;T) = q(Y,T)$, so that in particular $I_{q_{\piX}}(Y;T) = I_q(Y;T)$.
    \item $I_{q_{\piX}}(X;T) \leq I_q(X;T)$, where equality holds if and only if $q_{\piX}(T|X) = q(T|X)$. 
  \end{enumerate}
\end{lemma}

\begin{proof}
  $(i)$. For all $y \in \Ycal, t \in \Tcal$,
  \begin{align}
    q_{\piX}(y,t) &= \sum_{x \in \Xcal} p(y,x) q_{\piX}(t|x) = \sum_{j=1}^n \sum_{x \in \Xcal_j} p(y,x) q_{\piX}(t|x) \notag \\
    &= \sum_{j=1}^n \sum_{x, x' \in \Xcal_j} p(y,x) p(x') \frac{q(t|x')}{p(\Xcal_j)}  \notag \\
    &= \sum_{j=1}^n \sum_{x, x' \in \Xcal_j} p(y,x') p(x) \frac{q(t|x')}{p(\Xcal_j)} \label{eq:loc:ib:qYT_is_qprimeYT} \\
    &= \sum_{j=1}^n \sum_{x' \in \Xcal_j} p(y,x') q(t|x')  \notag \\
    &= \sum_{x \in \Xcal}  p(y,x') q(t|x') = q(y,t), \notag 
  \end{align}
  where the first and last lines use that $\{ \Xcal_j \}_j$ is a partition of $\Xcal$; and line \eqref{eq:loc:ib:qYT_is_qprimeYT} uses the definition of the sets $\Xcal_j$ through the equivalence relation $\sim_\Xcal$ (see equation \eqref{eq:def:equivalence_relation_ib}): i.e., for $x,x' \in \Xcal_j$, we have for all $y \in \Ycal$, $p(y|x) = p(y|x')$, which is equivalent to $p(y,x)p(x')=p(y,x')p(x)$.

  $(ii)$. We apply point $(ii)$ in Lemma \ref{lemma:factorisation_decreases_info}.
\end{proof}

But Lemma \ref{lemma:ib:relation_q_qprime} means that for the IB problem \eqref{eq:ib_problem_primal}, if we replace the chanel $q(T|X) \in C(\Xcal, \Tcal)$ by the corresponding $q_{\piX}(T|X)$, then $(i)$ the value for of the constraint function is unchanged, and $(ii)$ the value of the target function does not increase, with equality if and only if $q_{\piX}(T|X) = q(T|X)$. In particular, if $q(T|X)$ solves the IB problem, then we must have $q(T|X) = q_{\piX}(T|X)$: i.e., $q(T|X) = \qbar \circ {\piX}$.

\subsection{Explicit form of solutions for $\lambda = \Lambda$ (point $(i)$ in Theorem \ref{th:results_ib})}
\label{section:apx:solutions_ib_for_Lambda}

In this section, we prove point $(i)$ in Theorem \ref{th:results_ib}, i.e., that $\ib(\Lambda) = \left\{ \gamma \circ {\piX} \, : \ \ \gamma \in \Ccong(\Xcal_J, \Tcal) \right\}$.

We will here denote by $\Xcal_t^q$ the ``probabilistic pre-image'' $\Acal_t^q$ from Appendix \ref{section:apx:general_stuff} (see equation \eqref{eq:def:Scaltq}): i.e., for a channel $q = q(T|X)$ and all $t \in \Tcal$,
\begin{align} \label{eq:def:Xcaltq}
  \Xcal_t^q := \left\{ x \in \Xcal : \ \ q(t|x) > 0 \right\}.
\end{align}

\begin{lemma} \label{lemma:ib:charac_maximal_constraint}
  We have $I_q(T;Y) \leq \Lambda := I(X;Y)$, and the following are equivalent:
  \begin{enumerate}
    \item $I_q(T;Y) = \Lambda$.
    \item For all $t \in \supp(q(T))$, there exists a partition element $\Xcal_j \in \Xcalbar$ such that  $\Xcal_t^q \subseteq \Xcal_j$.
    \item The channel $\qbar(T|\Xcal_J) \in C(\Xcalbar, \Tcal)$ defined in \eqref{eq:def:qbar} is congruent.
  \end{enumerate}
\end{lemma}

The equivalence of $(i)$ and $(ii)$ means that the constraint $I_q(T;Y) = I(X;Y)$ holds if and only if $p(Y|x)$ is constant on the pre-image $\Xcal_t^q$ of every symbol $t$.

\begin{proof}
  $(i) \Leftrightarrow (ii)$. We have
  \begin{align*}
    I_q(T;Y) &= \sum_{y \in \supp(p(Y)), t \in \supp(q(T))} p(y) q(t|y) \log \left( \frac{q(t|y)}{q(t)} \right) \\ 
    &= \sum_{y \in \supp(p(Y)), t \in \supp(q(T))} p(y) \left( \sum_x p(x|y) q(t|x) \right) \log \left( \frac{\sum_x p(x|y) q(t|x)}{\sum_{x} p(x) q(t|x)}\right).
  \end{align*}
  But for all $y \in \supp(p(Y))$ and $t \in \supp(q(T))$, from the log-sum inequality \citep{csiszarInformationTheoryCoding2011}, with the convention $0\log(\frac{0}{0}) := 0$,
  \begin{align} \label{eq:loc:for_I_YT_leq_I_XY}
    \left( \sum_x p(x|y) q(t|x) \right) \log \left( \frac{\sum_x p(x|y) q(t|x)}{\sum_{x} p(x) q(t|x)} \right) \leq \sum_x p(x|y) q(t|x) \log \left( \frac{p(x|y) q(t|x)}{p(x) q(t|x)} \right).
  \end{align}
  So that, summing over $y$ and $t$, we get $I_q(Y;T) \leq I(X;Y)$, with equality if and only if for all $y \in \supp(p(Y)), t \in \supp(q(T))$, it holds in \eqref{eq:loc:for_I_YT_leq_I_XY}. From the equality case of the log-sum inequality \citep{csiszarInformationTheoryCoding2011}, the latter is equivalent to the existence of nonzero constants $(\alpha_{y,t})_{y \in \supp(p(Y)), t \in \supp(q(T))}$ such that
  \begin{align*}
    \forall x \in \Xcal, \quad p(x) q(t|x) = \alpha_{y,t} p(x|y) q(t|x),
  \end{align*}
  i.e., such that, for all $y \in \supp(p(Y)), t \in \supp(q(T))$, the quantity $\frac{p(x|y)}{p(x)}$ is constant on the subset of elements $x \in \Xcal$ for which $q(t|x)>0$. But the latter subset is precisely $\Xcal_t^q$ (see definition \eqref{eq:def:Scaltq}), and
  \begin{align*}
    \frac{p(x|y)}{p(x)} = \frac{1}{p(y)} p(y|x),
  \end{align*}
  where $\frac{1}{p(y)}$ does not depend on $x$. Thus we proved that $I(Y;T) = I(X;Y)$ holds if and only if for all $t \in \supp(q(T))$, the distribution $p(Y|x)$ does not depend on $x \in \Xcal_t^q$: i.e., if and only if for all $t \in \supp(q(T))$, there exists an $\Xcal_j$ such that  $\Xcal_t^q \subseteq \Xcal_j$.
  
  $(ii) \Leftrightarrow (iii)$. Apply Lemma \ref{lemma:general:charac_congruent}.
\end{proof}

Combining the previous results directly yields that
\begin{align*}
  \ib(\Lambda) \subseteq E := \left\{ \gamma \circ {\piX}, \quad \gamma \in \Ccong(\{1,\dots,n\}, \Tcal) \right\}.
\end{align*}
Indeed, fix a solution $q(T|X) \in \ib(\Lambda)$. Proposition \ref{prop:ib_factorises_for_all_lambda} proves that $q(T|X) = \qbar(T|\Xcal_J) \circ {\piX}$. But because we must have $I_q(Y;T) = \Lambda := I(X;Y)$, Lemma \ref{lemma:ib:charac_maximal_constraint} yields that $\qbar(T|\Xcal_J)$ is here congruent.

Let us now prove the converse inclusion, i.e., that $E\subseteq \ib(\Lambda)$.

\begin{lemma} \label{lemma:ib:value_target_constraint_for_solutions}
    For all $q(T|X) \in E$, we have $I_q(T;Y) = I(X;Y)$ and $I_{q}(X;T) = H({\piX}(X))$.
\end{lemma}
\begin{proof}
  From point $(i)$ in Lemma \ref{lemma:general:gamma_is_gammaq_and_I_is_HpiA}, $q(T|X) = \gamma \circ \piX$ implies $\gamma = \qbar(T|\Xcal_J)$. But by definition of $E$, here $\gamma$ is assumed congruent, thus $\qbar(T|\Xcal_J)$ is congruent. So that from Lemma \ref{lemma:ib:charac_maximal_constraint}, $I_q(T;Y) = I(X;Y)$. Point $(ii)$ in Lemma \ref{lemma:general:gamma_is_gammaq_and_I_is_HpiA} yields $I_{q}(X;T) = H({\piX}(X))$.
\end{proof}

Now, because the IB problem is defined as the minimisation of a continuous function on a compact domain, it has at least one solution, say $q_\ast(T|X)$, which we know belongs to $E$ from the the inclusion $\ib(\Lambda) \subseteq E$ that we already proved. But Lemma \ref{lemma:ib:value_target_constraint_for_solutions} then implies that for all $q(T|X) \in E$, we have $I_q(T;Y) = I_{q_\ast}(T;Y)$ and $I_{q}(X;T) = I_{q_\ast}(X;T)$. Thus any $q(T|X) \in E$ must also be a solution, i.e., $q(T|X) \in \ib(\Lambda)$. This ends the proof of point $(i)$ in Theorem \ref{th:results_ib}.

\subsection{End of the proof of Theorem \ref{th:results_ib}}
\label{section:apx:end_proof_results_ib}

Proposition \ref{prop:ib_factorises_for_all_lambda} ensures that for all $\lambda$ and $q(T|X) \in \ib(\lambda)$, we have the factorisation $q(T|X) = \qbar(T|\Xcal_J) \circ {\piX}$, with $\qbar(T|\Xcal_J)$ defined in \eqref{eq:def:qbar}. Thus, for all $\sigma \in \bij(\Xcal)$,
\begin{align} \label{eq:loc:proof_eib_characterises_invariances}
  \begin{split}
  \sigma \in \Gci \quad &\Leftrightarrow \quad \forall x \in \Xcal, \ \ \ p(Y|x) = p(Y | \sigma \cdot x) \\
  &\Leftrightarrow \quad \forall x \in \Xcal, \ \ \ x \sim_\Xcal \sigma \cdot x \\
  &\Leftrightarrow \quad \pi \circ \sigma = \pi \\
  &\Rightarrow \quad \qbar(T|\Xcal_J) \circ \pi \circ \sigma = \qbar(T|\Xcal_J) \circ \pi \\
  &\Leftrightarrow \quad q(T|X) \circ \sigma = q(T|X),
  \end{split}
\end{align}
which yields point $(iii)$ of Theorem \ref{th:results_ib}. 
Moreover, if we assume that $\lambda = \Lambda$, then from Lemma \ref{lemma:ib:charac_maximal_constraint}, here $\qbar(T|\Xcal_J)$ is a congruent channel, i.e., there exists a function $f$ such that $f \circ \qbar(T|\Xcal_J)$ is the identity on $\Xcalbar$. Therefore the only implication in \eqref{eq:loc:proof_eib_characterises_invariances} becomes an equivalence as well, which yields point $(ii)$ of Theorem \ref{th:results_ib}.

Let us now prove point $(iv)$ in Theorem \ref{th:results_ib}. The statement is equivalent to proving that the equivalence relation defined by the partition in orbits under $\Gci$, which we denote here by $\sim_{\ci}$, coincides with the equivalence relation $\sim_{\Xcal}$ defined in \eqref{eq:def:equivalence_relation_ib}. Moreover, by definition of an orbit, $x \sim_{\ci} x'$ means that there exitst $\sigma \in \bij(\Xcal)$ such that $(i)$ $\sigma \in \Gci$, i.e., $p(Y|\sigma \cdot x'') = p(Y|x'')$ for all $x'' \in \Xcal$, and $(ii)$ $x' = \sigma \cdot x$. 

Thus $x \sim_{\ci} x'$ clearly implies $p(Y|x) = p(Y|x')$, i.e., $x \sim_{\Xcal} x'$. Conversely, let us fix $x,x'\in \Xcal$ such that $x \sim_{\Xcal} x'$. We define $\sigma$ as the transposition that permutes $x$ and $x'$, and fixes all the other elements of $\Xcal$. It is straightforward to verify that
$\sigma$ satisfies points $(i)$ and $(ii)$ above, i.e., that we have $x \sim_{\ci} x'$.

\section{Appendix for Section \ref{section:sib}}
\label{section:apx:sib}

\subsection{On the projection on the exponential family}
\label{section:apx:sib_on_projection_exponential}

Let us recall that $\cl \Ecal$ denotes the topological closure of the exponential family $\Ecal$. Here, we denote by $\ptilde \in \cl \Ecal$ the unique distribution \citep{ayInformationGeometry2017} which achieves the minimum in $\inf_{r \in \cl \Ecal} D(p||r)$; but we do not assume, a priori, that $\ptilde$ minimises $\inf_{r \in \cl \Ecal} D(\kappa \cdot p||\kappa \cdot r)$. Note that we always have $\supp(p) \subseteq \supp(\ptilde)$, because otherwise $D(p||\ptilde) = +\infty > \inf_{r \in \Ecal} D(p||r)$. In particular, whenever $p(A)$ is full support, then $\ptilde(A)$ is full support as well. In the latter case, $\ptilde$ is thus both in $\cl \Ecal$ and the interior of the simplex $\Delta_\Acal$, which implies that $\ptilde \in \Ecal$.

Let us now prove that $\ptilde$ also achieves the latent space divergence, i.e., for all $\kappa \in C(\Acal, \Tcal)$, we automatically have $D(p||\ptilde) = \inf_{r \in \cl \Ecal} D(\kappa \cdot p || \kappa \cdot r)$. Indeed, for all $r \in \cl \Ecal$ and with the convention $0\log(\frac{0}{0}) := 0$,
\begin{align}
    D(\kappa \cdot p||\kappa \cdot \ptilde) - D(\kappa \cdot p||\kappa \cdot r) &= \sum_{t \in \Tcal} (\kappa \cdot p)(t) \log \left( \frac{(\kappa \cdot r)(t)}{(\kappa \cdot \ptilde)(t)} \right) \notag \\
    &= \sum_{t \in \Tcal} \sum_{a \in \Acal} p(a) q(t|a) \log \left( \frac{\sum_{a \in \Acal} r(a) q(t|a)}{\sum_{a \in \Acal} \ptilde(a) q(t|a)} \right) \notag \\
    &\leq \sum_{t \in \Tcal} \sum_{a \in \Acal} p(a) q(t|a) \log \left( \frac{r(a) q(t|a)}{\ptilde(a) q(t|a)} \right) \label{eq:loc:reparametrisation_sib_1} \\
    &= \sum_{a \in \Acal} p(a) \log \left( \frac{r(a)}{\ptilde(a)} \right) \notag \\
    &= D(p||\ptilde) - D(p||r) \notag \\
    &\leq 0, \label{eq:loc:reparametrisation_sib_2}
\end{align}
where line \eqref{eq:loc:reparametrisation_sib_1} uses the log-sum inequality \citep{csiszarInformationTheoryCoding2011}, and line \eqref{eq:loc:reparametrisation_sib_2} that $D(p||\ptilde) = \inf_{r \in \cl \Ecal} D(p||r)$. In particular, $\ptilde$ is the unique distribution in $\cl \Ecal$ minimising both $\inf_{r \in \cl \Ecal} D(p||r)$ and $\inf_{r \in \cl \Ecal} D(\kappa \cdot p||\kappa \cdot r)$.

\subsection{Proof of Theorem \ref{th:results_sib}}
\label{section:apx:proofs_sib}

We will use the notations and definitions from Section \ref{section:sib_general} and Appendix \ref{section:apx:general_stuff}, which are consistent. We also write $q(T,A)$, $q_\pi(A,T)$, $\qtilde(T,A)$ $\qtilde_\pi(T,A)$ the joint distributions defined resp. by $q(t,a) := p(a) q(t|a)$, $q_\pi(a,t) := p(a) q_\pi(t|a)$, $\qtilde(t,a) = \ptilde(a) q(t|a)$ and $\qtilde_\pi(t,a) = \ptilde(a) q_\pi(t|a)$. Note that for $\kappa = q(T|A)$, the quantity $D(\kappa \cdot p || \kappa \cdot \ptilde)$ then becomes $D(q(T)||\qtilde(T))$, which will be a more convenient notation in the proofs below.

\subsubsection{Factorisation for all parameter $\lambda$}
\label{section:apx:proof_sib_factorises_for_all_lambda}

\begin{proposition} \label{prop:sib_factorises_for_all_lambda}
  For all $0 \leq \lambda \leq \Lambda$, every solution $q(T|A) \in \iname(\lambda)$ factorises as $q(T|A) = \qbar(T|\Acal_J) \circ \pi$.
\end{proposition}

The proof will derive from the following lemma:

\begin{lemma} \label{lemma:sib:relation_q_qprime}
  For every $q(T|A) \in C(\Acal, \Tcal)$, we have:
  \begin{enumerate}
    \item $q_{\pi}(T) = q(T)$ and $\qtilde_\pi(T) = \qtilde(T)$. In particular, $D(q_\pi(T)||\qtilde_\pi(T)) = D(q(T)||\qtilde(T))$.
    \item $I_{q_{\pi}}(A;T) \leq I_q(A;T)$, where equality holds if and only if $q_{\pi}(T|A) = q(T|A)$. 
  \end{enumerate}
\end{lemma}

\begin{proof}
  $(i)$. $q_{\pi}(T) = q(T)$ is point $(i)$ in Lemma \ref{lemma:factorisation_decreases_info}. Moreover, for all $t \in \Tcal$,
  \begin{align}
    \qtilde_\pi(t) &= \sum_{a \in \Acal} q_\pi(t|a) \ptilde(a) \notag \\
    &= \sum_j \sum_{a \in \Acal_j} q_\pi(t|a) \ptilde(a) = \sum_j \sum_{a \in \Acal_j} \frac{q(t,\Acal_j)}{p(\Acal_j)} \ptilde(a) \notag\\
    &= \sum_j \sum_{a, a' \in \Acal_j} \frac{1}{p(\Acal_j)} q(t|a') p(a') \ptilde(a) \notag \\
    &= \sum_j \sum_{a, a' \in \Acal_j} \frac{1}{p(\Acal_j)} q(t|a') p(a) \ptilde(a') \label{eq:local:tqprime_is_tq_2} \\
    &= \sum_j \sum_{a' \in \Acal_j}  q(t|a') \ptilde(a') = \sum_{a' \in \Acal}  q(t|a') \ptilde(a') \notag\\
    &= \qtilde(t) \notag
  \end{align}
  where \eqref{eq:local:tqprime_is_tq_2} uses the definition of the sets $\Acal_j$ through the equivalence relation $\sim$ defined in \eqref{eq:def:equivalence_relation_sib}, i.e., $p(a) \ptilde(a') = p(a) \ptilde(a')$ for $a,a' \in \Acal_j$.

  $(ii)$. We apply point $(ii)$ in Lemma \ref{lemma:factorisation_decreases_info}.
\end{proof}

But Lemma \ref{lemma:sib:relation_q_qprime} means that for the DIB problem \eqref{eq:def:sib}, if we replace the channel $q(T|A)$ by the corresponding $q_{\pi}(T|A)$, then $(i)$ the value for of the constraint function is unchanged, and $(ii)$ the value of the target function does not increase, with equality if and only if $q_{\pi}(T|A) = q(T|A)$. In particular, if $q(T|A)$ solves the DIB problem, then we must have $q(T|A) = q_{\pi}(T|A)$: i.e., $q(T|A) = \qbar(T|\Acal_J) \circ {\pi}$.

\subsubsection{Explicit form of solutions for $\lambda = \Lambda$ (Theorem \ref{th:results_sib})}
\label{section:apx:solutions_sib_for_Lambda}

In this section, we prove Theorem \ref{th:results_sib}, i.e., that $\iname(\Lambda) = \left\{ \gamma \circ {\pi} \, : \ \ \gamma \in \Ccong(\Acalbar, \Tcal) \right\}$. Recall that $\Acal_t^q$ is the ``probabilistic pre-image of $t$ through $q(T|A)$'' (see equation \eqref{eq:def:Scaltq}).

\begin{lemma} \label{lemma:sib:charac_maximal_constraint}
  We have $D(q(T)||\qtilde(T)) \leq \Lambda := D(p(A)||\ptilde(A))$, and the following are equivalent:
  \begin{enumerate}
    \item $D(q(T)||\qtilde(T)) = \Lambda$.
    \item For all $t \in \supp(q(T))$, there exists a partition element $\Acal_j \in \Acalbar$ such that  $\Acal_t^q \subseteq \Acal_j$.
    \item The channel $\qbar(T|\Acal_J) \in C(\Acalbar, \Tcal)$ defined in \eqref{eq:def:qbar} is congruent.
  \end{enumerate}
\end{lemma}

\begin{proof}
  $(i) \Leftrightarrow (ii)$. We have
  \begin{align*}
    D(q(T)||\qtilde(T)) &= \sum_{t \in \supp(q(T))} \left(\sum_{a \in \mathcal{S}} q(t|a) p(a) \right) \log\left( \frac{\sum_{a \in \mathcal{S}} q(t|a) p(a)}{\sum_{a \in \mathcal{S}} q(t|a) \ptilde(a)} \right),
  \end{align*}
  
  But for all $t \in \supp(q(T))$, from the log-sum inequality \citep{csiszarInformationTheoryCoding2011}, with the convention $0\log(\frac{0}{0}) := 0$,
  \begin{align} \label{eq:loc:for_Dqqprime_leq_Dpprime}
    \left(\sum_{a \in \mathcal{S}} q(t|a) p(a) \right) \log\left( \frac{\sum_{a \in \mathcal{S}} q(t|a) p(a)}{\sum_{a \in \mathcal{S}} q(t|a) \ptilde(a)} \right) \leq \sum_{a \in \mathcal{S}} q(t|a) p(a)  \log\left( \frac{q(t|a) p(a)}{q(t|a) \ptilde(a)} \right).
  \end{align}

  So that, summing over $t$, we get $ D(q(T)|| \qtilde(T)) \leq D(p(A)||\ptilde(A))$, with equality if and only if for all $t \in \Tcal$, it holds in \eqref{eq:loc:for_Dqqprime_leq_Dpprime}. From the equality case of the log-sum inequality \citep{csiszarInformationTheoryCoding2011}, the latter is equivalent to the existence of nonzero constants $(\alpha_{t})_{t \in \supp(q(T))}$ such that
  \begin{align*}
    \forall a \in \Acal \quad q(t|a) p(a) = \alpha_{t} q(t|a) \ptilde(a),
  \end{align*}
  i.e., such that, for all $t \in \supp(q(T))$, we have $p(a) = \alpha_{t} \ptilde(a)$ for all $a \in \Acal$ such that $q(t|a)>0$, i.e.,
  \begin{align*} 
    \forall t \in \supp(q(T)), \ \forall a \in \Acal_t^q, \ \frac{p(a)}{\ptilde(a)} = \alpha_{t}
  \end{align*}
  In the above, note that the fraction $\frac{p(a)}{\ptilde(a)}$ does make sense, because here  $\supp(\ptilde) = \Acal$ (see Section \ref{section:apx:sib_on_projection_exponential}). Thus we proved that equality holds in \eqref{eq:loc:for_Dqqprime_leq_Dpprime} if and only if for all $t \in \supp(q(T))$, the quotient $\nicefrac{p(a)}{\ptilde(a)}$ is constant on $\Acal_t^q$, i.e., if and only if for all $t \in \supp(q(T))$, there exists an $\Acal_j$ such that  $\Acal_t^q \subseteq \Acal_j$.
  
  $(ii) \Leftrightarrow (iii)$. Apply Lemma \ref{lemma:general:charac_congruent}.
\end{proof}

Combining the previous results directly yields that
\begin{align*}
  \iname(\Lambda) \subseteq E := \left\{ \gamma \circ {\pi}, \quad \gamma \in \Ccong(\Acalbar, \Tcal) \right\}.
\end{align*}
Indeed, fix a solution $q(T|A) \in \iname(\Lambda)$. Proposition \ref{prop:sib_factorises_for_all_lambda} proves that $q(T|A) = \qbar(T|\Acal_J) \circ {\pi}$. But because we must have $D(q(T)||\qtilde(T)) = \Lambda := D(p(A)||\ptilde(A))$, Lemma \ref{lemma:sib:charac_maximal_constraint} yields that $\qbar(T|\Acal_J)$ is here congruent.

Let us now prove the converse inclusion, i.e., that $E\subseteq \iname(\Lambda)$.

\begin{lemma} \label{lemma:sib:value_target_constraint_for_solutions}
    For all $q(T|A) \in E$, we have $D(q(T)||\qtilde(T)) =  D(p(A)||\ptilde(A))$ and $I_{q}(A;T) = H({\pi}(A))$.
\end{lemma}
\begin{proof}
  From point $(i)$ in Lemma \ref{lemma:general:gamma_is_gammaq_and_I_is_HpiA}, $q(T|A) = \gamma \circ \pi$ implies $\gamma = \qbar(T|\Acal_J)$. But by definition of $E$, here $\gamma$ is assumed congruent, thus so is $\qbar(T|\Acal_J)$. So that from Lemma \ref{lemma:sib:charac_maximal_constraint}, $D(q(T)||\qtilde(T)) = \Lambda = D(p(A)||\ptilde(A))$. Point $(ii)$ in Lemma \ref{lemma:general:gamma_is_gammaq_and_I_is_HpiA} yields $I_{q}(A;T) = H({\pi}(A))$.
\end{proof}

Now, because the DIB problem is defined as the minimisation of a continuous function on a compact domain, it has at least one solution, say $q_\ast(T|A)$, which we know belongs to $E$ from the the inclusion $\iname(\Lambda) \subseteq E$ that we already proved. But Lemma \ref{lemma:sib:value_target_constraint_for_solutions} then implies that for all $q(T|A) \in E$, we have $D(q(T)||\qtilde(T)) = D(q_\ast(T)||\qtilde_\ast(T))$ and $I_{q}(A;T) = I_{q_\ast}(A;T)$. Thus any $q(T|A) \in E$ must also be a solution, i.e., $q(T|A) \in \iname(\Lambda)$. This ends the proof of point $(i)$ in Theorem \ref{th:results_sib}.

\subsection{The Divergence IB captures equivariances (proof of Theorem \ref{th:charac_equivariances_with_sib})}
\label{section:apx:proof_charac_equivariances_with_sib}

$(i)$. The proof is almost identical to that of point $(ii)$ of Theorem \ref{th:results_ib} (see Appendix \ref{section:apx:end_proof_results_ib}). Here $\inameop(\Lambda)$ denotes the solutions to the specific $\iname$ problem defined in Section \ref{section:sib_applied_to_equivariances}, and $\pi$ the corresponding projection defined by 
\begin{align*}
  (x,y) \sim (x',y') \quad \Leftrightarrow \quad p(x,y) p(x') = p(x',y') p(x).
\end{align*}
Proposition \ref{prop:sib_factorises_for_all_lambda} ensures that for all $\lambda$ and $\kappa = q(T|A) \in \iname_\ce(\lambda)$, we have the factorisation $\kappa = \bar{\kappa} \circ \pi$, with $\bar{\kappa} = \qbar(T|\Scal_J)$ defined in \eqref{eq:def:qbar}. Thus, for all $(\sigma, \tau) \in \bij(\Xcal) \times \bij(\Ycal)$,
\begin{align} \label{eq:loc:proof_sib_characterises_equivariances}
  \begin{split}
  (\sigma, \tau) \in \Gce \quad &\Leftrightarrow \quad \forall (x,y) \in \Xcal \times \Ycal, \ \ \ p(y|x) = p(\tau \cdot y | \sigma \cdot x) \\
  &\Leftrightarrow \quad \forall (x,y) \in \Xcal \times \Ycal, \ \ \ p(x,y) p(\sigma \cdot x) = p(\sigma \cdot x, \tau \cdot y) p(x) \\
  &\Leftrightarrow \quad \forall (x,y) \in \Xcal \times \Ycal, \ \ \ (x,y) \sim (\sigma \cdot x, \tau \cdot y) \\
  &\Leftrightarrow \quad \pi \circ (\sigma, \tau) = \pi \\
  &\Rightarrow \quad \bar{\kappa} \circ \pi \circ (\sigma, \tau) = \bar{\kappa} \circ \pi \\
  &\Leftrightarrow \quad \kappa \circ (\sigma, \tau) = \kappa,
  \end{split}
\end{align}
where the first equivalence follows easily from the definition of equivariance (see Lemma 15 in \citep{charvinInformationTheoryBasedDiscovery2023a} for details). 
This yields point $(ii)$. 
Moreover, if we assume that $\lambda = \Lambda$, then from Lemma \ref{lemma:sib:charac_maximal_constraint}, here $\bar{\kappa}$ is a congruent channel, i.e., there exists a function $f$ such that $f \circ \bar{\kappa}$ is the identity on $\Acalbar$. Thus the only implication in \eqref{eq:loc:proof_sib_characterises_equivariances} becomes an equivalence as well, which yields point $(i)$ of Theorem \ref{th:charac_equivariances_with_sib}.

$(iii)$. Here, the reasoning used for the proof of point $(iii)$ in Theorem \ref{th:results_ib} does not work. Indeed the transposition $\Phi \in \bij(\Xcal \times \Ycal)$ that permutes two pairs $(x,y)$ and $(x',y')$ and fixes all the other ones does not have a split form $\sigma \otimes \tau$ for some $(\sigma, \tau) \in \bij(\Xcal) \times \bij(\Ycal)$.

Moreover, let $\Xcal = \{1,2,3\}$ and $\Ycal = \{1,2\}$, with $p(X)$ uniform and $p(Y|X)$ defined through the row transition matrix
\begin{align*}
  \begin{pmatrix}
    c & p_{12} \\
    p_{21} & c \\
    p_{31} & p_{32}
  \end{pmatrix}
\end{align*}
where we choose $c, p_{12}$, $p_{21}$, $p_{31}$, and $p_{32}$ pairwise distinct. It can be easily shown that this channel has no non-trivial equivariances, i.e., $\Gce = \{e_{\Xcal \times \Ycal}\}$, so that the projection on orbits $\pice$ is the identity of $\Xcal \times \Ycal$. Yet the projection $\pi$ defined by the the relation $\sim$ will here identify the two pairs $(x,y)$ such that $p(y|x)=c$. Therefore $\pice \neq \pi$.

\subsection{Relation to the Intertwining IB}
\label{section:apx:relation_to_iib}

Our work is heavily inspired from that in \citep{charvinInformationTheoryBasedDiscovery2023a}; in this section we explicitly relate the two. The latter reference considered the Intertwining IB problem, namely,
\begin{align} \label{eq:iib_def}
  \iib(\lambda) \ := \ \argmin_{\substack{\kappa \, \in \, C(\Xcal \times \Ycal, \Tcal) \; : \\ \ D(\kappa \cdot p(X,Y)||\kappa \cdot p(X)p(Y)) \geq \lambda}} \; I_\kappa(X,Y;T).
\end{align}
It can readily be verified this is a DIB problem with $\Ecal =  \Delta_\Xcal \otimes \Delta_\Ycal$ and $C = C(\Xcal \times \Ycal, \Tcal)$.

Problem \eqref{eq:iib_def} is used in \citep{charvinInformationTheoryBasedDiscovery2023a} to characterise equivariances under specific conditions: if $(i)$ the distribution $p(X,Y)$ is discrete and full support, and $(ii)$ $p(Y)$ is uniform, then the solution $\kappa$ to \eqref{eq:ib_problem_primal_iib_formulation} with $\lambda = I(X;Y)$ are such that a pair $(\sigma, \tau) \in \bij(\Xcal) \times \bij(\Ycal)$ is an equivariance if and only if $\kappa \circ (\sigma \otimes \tau) = \kappa$.

Thus, Section \ref{section:sib_applied_to_equivariances} in this work is an improvement on the latter result: here, we replace the Intertwining IB problem by the similar problem $\iname_\ce$, from which we obtain the same characterisation as above, except that the assumption $(ii)$ that $p(Y) = \Ucal(\Ycal)$ can now be dropped. Note that the latter assumption holds precisely when the hierarchical model $\Ecal_{\text{IIB}} := \Delta_\Xcal \otimes \Delta_\Ycal$ from \citep{charvinInformationTheoryBasedDiscovery2023a} coincides with the hierachical model $\Ecal_\ce := \Delta_\Xcal \otimes \{\Ucal(\Ycal)\}$ from the present article. In this sense, we have clarified that $\Ecal_\ce$, and not $\Ecal_{\text{IIB}}$, is the ``correct'' hierarchical model to characterise --- and soften --- equivariances. 

More broadly, the present work generalises the Intertwining IB framework to arbitrary (discrete) underlying space $\Acal$ and arbitrary hierarchichal model $\Ecal$. Note that our definition of soft equivariances is directly analogous to that proposed in \citep{charvinInformationTheoryBasedDiscovery2023a}, even though it is different, as it based on a DIB problem defined by $\Ecal_\ce $ and not $\Ecal_{\text{IIB}}$.

Eventually, whereas \citep{charvinInformationTheoryBasedDiscovery2023a} does not provide algorithms to compute solutions to the IIB problem \eqref{eq:iib_def}, here we design a BA algorithm for a general $\iname$ problem (without constraints on encoder shape) with the same convergence guarantees as the BA for the classic IB (see Section \ref{section:apx:blahut_arimoto}), which we use for synthetic numerical experiments in Section \ref{section:numerical_experiments}.

\subsection{The classic IB is a Divergence IB}
\label{section:apx:ib_is_sib}

Ref. \citep{charvinInformationTheoryBasedDiscovery2023a} proves that the classic IB can be formulated as an Intertwining IB with specific constraints on the shape of compression channels. More precisely, define $\Tcal$ as $\Tcal := \Tcal_{\text{IB}} \times \Ycal$ with $\Tcal_{\text{IB}} := \mathbb{N}$, \footnote{This choice is formally equivalent to $\Tcal := \mathbb{N}$, as there is a bijection between $\mathbb{N}$ and $\mathbb{N} \times \Ycal$.} and consider the set
\begin{align*}
  C_{\text{IB}(X,Y)} := \{ \kappa_{\mathcal{X}} \otimes e_{\mathcal{Y}} \, : \ \;  \kappa_{\mathcal{X}} \in  C(\mathcal{X}, \mathcal{T}_{\text{IB}}) \} \subset C(\Xcal \times \Ycal, \Tcal_{\text{IB}} \times \Ycal)
\end{align*}
of channels that can compress the $\mathcal{X}$ coordinate but copy the $\mathcal{Y}$ coordinate. This leads to the problem
\begin{align} \label{eq:ib_problem_primal_iib_formulation}
  \iib_{C_{\text{IB}}}(\lambda) \ := \ \argmin_{\substack{\kappa \, \in \, C_{\text{IB}(X,Y)} \; : \\ \ D(\kappa(p(X,Y))||\kappa(p(X)p(Y))) = \lambda}} \; I_\kappa(X,Y;T),
\end{align}
Then:
\begin{proposition}[\citep{charvinInformationTheoryBasedDiscovery2023a}, Prop. 5]
  \label{prop:ib_equivalent_to_iib_with_constraint}
    For every $0 \leq \lambda \leq I(X;Y)$, a channel $\kappa_{\mathcal{X}} \otimes e_{\mathcal{Y}} \in C_{\ib}(X,Y)$ solves the problem \eqref{eq:ib_problem_primal_iib_formulation} if and only if $\kappa_{\mathcal{X}} = \kappa_{\mathcal{X}}(T_{\ib}|X)$ solves the IB problem \eqref{eq:ib_problem_primal}.
\end{proposition}
In this sense, the classic IB is equivalent to
the problem \eqref{eq:ib_problem_primal_iib_formulation}. Importantly, in can be easily verified that the latter is a DIB with $C = C_{\text{IB}(X,Y)}$ and $\Ecal = \Delta_\Xcal \otimes \Delta_\Ycal$.

However, for the sake of consistency with the results presented in this work, let us also prove that the classic IB is equivalent to a DIB with still $C = C_{\text{IB}(X,Y)}$, but now 
\begin{align*}
  \Ecal = \Ecal_\ce := \{ r(X)\Ucal(\Ycal), \ r(X) \in \Delta_{\Xcal} \},
\end{align*}
which is the exponential family used in Section \ref{section:sib_applied_to_equivariances} to fully characterise channel equivariances.

As mentioned in Section \ref{section:sib_applied_to_equivariances}, we have $D(p(X,Y)||\Ecal_\ce) = D(p(X,Y)||p(X)\Ucal(\Ycal))$. Moreover for $\kappa = \kappa_\Xcal \otimes e_\Ycal \in C_{\text{IB}}$, we have $\kappa \cdot p(X,Y) = q(T,Y)$, while $\kappa \cdot p(X)\Ucal(\Ycal) = q(T)\Ucal(\Ycal)$ and $\kappa \cdot p(X)p(Y) = q(T)p(Y)$, where the joint distribution $q(T,X,Y)$ is defined using $\kappa = q(T|X)$ together with $p(X,Y)$ and the Markov chain $T-X-Y$. Thus
\begin{align*}
  D(\kappa \cdot p(X,Y) || \kappa \cdot \Ecal_\ce) &= D(q(T,Y)||q(T)\Ucal(\Ycal)) \\
  &=  D(q(T,Y)||q(T)p(Y)) + D(p(Y)||\Ucal(\Ycal)) \\
  &=  D(\kappa(p(X,Y))||\kappa(p(X)p(Y)))  + D(p(Y)||\Ucal(\Ycal)).
\end{align*}
Therefore the constraint function in the DIB defined in \eqref{eq:ib_problem_primal_iib_formulation}, and the constraint function for the same problem but with $\Ecal_{\text{IIB}}$ replaced by $\Ecal_\ce$, differ by a constant $K$ that depends only on $p(Y)$, which is here fixed. In particular, the corresponding DIB problems are equivalent, in that for all $0 \leq \lambda \leq D(p(X,Y)||p(X)p(Y))$,
\begin{align*}
  \iib_{C_{\text{IB}}}(\lambda) = \iname_{\Ecal_\ce, C_{\text{IB}}}(\lambda + K).
\end{align*}
As Proposition~\ref{prop:ib_equivalent_to_iib_with_constraint} proves that $\iib_{C_{\text{IB}}}$ is equivalent to the classic IB, this proves that $\iname_{\Ecal_\ce, C_{\text{IB}}}$ is also equivalent to the classic IB (up to shifting the trade-off parameter $\lambda$ by a constant $K$).

In particular, our framework captures channel invariances --- which are a special case of channel equivariances with trivial action on the output space --- by using the exponential family $\Ecal_\ce$ that captures equivariances, and imposing the additional constraint $C_{\text{IB}}$ of only compressing the input space but leaving the output space unchanged.

\subsection{The Divergence IB captures distribution invariances (proof of Theorem \ref{th:charac_distrib_invariances_with_sib})}
\label{section:apx:proof_charac_distrib_invariances_with_sib}

The proof is almost identical to that of points $(ii)$ and $(iii)$ of Theorem \ref{th:results_ib} (see Appendix \ref{section:apx:end_proof_results_ib}). Here $\inameop(\Lambda)$ denotes the solutions to the specific $\iname$ problem defined in Section \ref{section:sib_applied_to_distrib_invariances}, and $\pi$ the corresponding projection defined by
\begin{align} \label{eq:def:equivalence_relation_sib_distrib_invariances_apx}
  a \sim a \quad \Leftrightarrow \quad p(a) = p(a')
\end{align}
Proposition \ref{prop:sib_factorises_for_all_lambda} ensures that for all $\lambda$ and $\kappa \in \inameop(\lambda)$, we have the factorisation $\kappa = \bar{\kappa} \circ \pi$, with $\bar{\kappa}$ defined in \eqref{eq:def:qbar}. Thus, for all $\Phi \in \bij(\Acal)$,
\begin{align} \label{eq:loc:proof_eib_characterises_distrib_invariances}
  \begin{split}
  \Phi \in \Gdi \quad &\Leftrightarrow \quad \forall a \in \Acal, \ \ \ p(a) = p(\Phi \cdot a) \\
  &\Leftrightarrow \quad \forall a \in \Acal, \ \ \ a \sim \Phi \cdot a \\
  &\Leftrightarrow \quad \pi \circ \Phi = \pi \\
  &\Rightarrow \quad \bar{\kappa} \circ \pi \circ \Phi = \bar{\kappa} \circ \pi \\
  &\Leftrightarrow \quad \kappa \circ \Phi = \kappa.
  \end{split}
\end{align}
This yields point $(iii)$ of Theorem \ref{th:results_sib}. 
Moreover, if we assume that $\lambda = \Lambda$, then from Lemma \ref{lemma:sib:charac_maximal_constraint}, here $\bar{\kappa}$ is a congruent channel, i.e., there exists a function $f$ such that $f \circ \bar{\kappa}$ is the identity on $\Scalbar$. Therefore the only implication in \eqref{eq:loc:proof_eib_characterises_distrib_invariances} becomes an equivalence as well, which yields point $(i)$.

$(iii)$. The statement is equivalent to proving that the equivalence relation defined by the partition in orbits under $\Gdi$, which we denote here by $\sim_{\di}$, coincides with the equivalence relation $\sim$ defined in \eqref{eq:def:equivalence_relation_sib_distrib_invariances_apx}. Moreover, by definition of an orbit, $a \sim_{\di} a'$ means that there exitst $\Phi \in \Gdi$ such that $(i)$ $\Phi \in \Gdi$, i.e., $p(\Phi \cdot a'') = p(a'')$ for all $a'' \in \Acal$, and $(ii)$ $a' = \Phi \cdot a$. 

Thus $a \sim_{\di} a'$ clearly implies $p(a) = p(a')$, i.e., $a \sim a'$. Conversely, let us fix $a,a'\in \Acal$ such that $a \sim a'$. We define $\Phi \in \bij(\Acal)$ as the transposition that permutes $a$ and $a'$, and fixes all the other elements of $\Acal$. It is straightforward to verify that
$\Phi$ satisfies points $(i)$ and $(ii)$ above, i.e., that we have $a \sim_{\di} a'$.

\section{Appendix for section \ref{section:sib_tools}}
\label{section:apx:tools}

In this appendix, the distribution $p(A)$ is allowed to not be full support, and we denote by $\Scal$ this support. In this case, there is still a unique distribution $\ptilde$ in the closure $\cl \Ecal$ of $\Ecal$ such that $D(p||\ptilde) = \inf_{r \in \Ecal} D(p||\ptilde)$ \citep{ayInformationGeometry2017}. We denote by $\Scaltilde$ the support of $\ptilde$. Note that $D(p||\ptilde) = \inf_{r \in \Ecal} D(p||r) < +\infty$ implies $\Scal \subseteq \Scaltilde$. We also assume now that $\Tcal$ is finite, and we define the DIB Lagrangian, on $C(\Acal, \Tcal)$, as
\begin{align} \label{eq:def:sib_lagrangian_apx}
  \mathcal{L}_{\beta}(q(T|A)) :=  I_q(A;T) - \beta D(q(T)||\tilde{q}(T)).
\end{align}
After technical preliminaries in Section \ref{section:apx:blahut_arimoto_extend_from_support}, we prove in Section \ref{section:apx:blahut_arimoto_self_consistent} the following necessary condition for local minimisers $\kappa = q(T|A)$ of \eqref{eq:def:sib_lagrangian_apx}: for all $a \in \supp(p(A))$ and $t \in \supp(q(T))$,
\begin{align}   \label{eq:self_consistent_equation}
  q(t|a) = \frac{1}{Z(a,\beta)} q(t) \exp\left[ - \beta \left( \frac{q(t)\ptilde(a)}{\tilde{q}(t)p(a)} - \log\left( \frac{q(t) \ptilde(a)}{\tilde{q}(t) p(a)} \right) - 1 \right)\right],
\end{align}
where $q(t) := \sum_a p(a) q(t|a)$ and $\qtilde(t) := \sum_a \ptilde(a) q(t|a)$, with $Z(a, \beta)$ a normaliser. From this fixed-point equation, we obtain a Blahut-Arimoto (BA) algorithm with the same guarantees as BA for the classic IB \citep{tishbyInformationBottleneckMethod2000} (see Section \ref{section:apx:blahut_arimoto_algorithm}). Eventually, we provide more details on effective cardinality in Section \ref{section:apx:charac_effcard_with_frac}.

\subsection{Minimisers on $\Scal \subseteq \Acal$ yield minimisers on $\Acal$}
\label{section:apx:blahut_arimoto_extend_from_support}

In this section, we reduce the minimisation of $\Lcal_\beta$ on $C(\Acal, \Tcal)$ to a minimisation over channels defined only on the support $\Scal$ of $p = p(A)$. More precisely, we show that a minimiser of $\Lcal_\beta$ can always be obtained the following way: choose a minimiser $\kappa \in C(\Scal, \Tcal)$ of the Lagrangian $\Lcal_\beta$ restricted to $C(\Scal,\Tcal)$, and extend it to a channel in $C(\Acal,\Tcal)$ by sending $\Acal \setminus \Scal$ on a dummy symbol $t_0 \notin \supp(\kappa \cdot p)$. This allows us, in our numerical experiments, to reduce the computation of minimisers of \eqref{eq:def:sib_lagrangian_apx} to that of the same Lagrangian restricted to channels defined only on the support $\Scal \subseteq \Acal$. This restriction to the support allows us to then use the BA algorithm (see Section \ref{section:apx:blahut_arimoto}), which indeed can only be applied on the support $\Scal$.

Let $q(T|A) \in C(\Acal, \Tcal)$. We write $q_\Scal(T|A) \in C(\Scal, \Tcal)$ and $p_\Scal(A)$ the restrictions of $q(T|A)$, resp. $p(A)$, to $\Scal \subseteq \Acal$: i.e., $q_\Scal(t|a) := q(t|a)$ and $p_\Scal(a) := p(a)$ for all $a \in \Scal$, $t \in \Tcal$ --- note that these are abuses of notation, as the input alphabet of $q_\Scal(T|A)$ is actually only $\Scal$, and similarly $p_\Scal(A)$ is only defined on $\Scal$. Of course $p_\Scal$ is a probability on $\Scal := \supp(p(A))$. We extend all the notations relating to $q(T|A)$ in Section \ref{section:apx:sib} to $q_\Scal(T|A)$; in particular, for $a \in \Scal, t \in \Tcal$,
\begin{align}
  q_\Scal(t) &:= \sum_{a \in \Scal} q_\Scal(t,a) := \sum_{a \in \Scal} p_\Scal(a) q_\Scal(t|a) = \sum_{a \in \Scal} p(a) q(t|a) = q(t), \label{eq:loc:def_q_S_T} \\
  \qtilde_\Scal(t) &:= \sum_{a \in \Scal} \qtilde_\Scal(t,a) := \sum_{a \in \Scal} \ptilde_\Scal(a) q_\Scal(t|a) := \sum_{a \in \Scal} \ptilde(a) q(t|a), \notag
\end{align}
or
\begin{align}  
  \Lcal_{\beta, \Scal} (q_\Scal(T|A)) &:= I_{q_\Scal}(A;T) - \beta D(q_\Scal(T)||\qtilde_\Scal(T)) \label{eq:def:sib_lagrangian_apx_on_S} \\
  &:= \sum_{a \in \Scal, t \in \supp(q_\Scal(T))} p_\Scal(a) q_\Scal(t|a) \log\left( \frac{q_\Scal(t|a)}{q_\Scal(t)} \right) - \beta  \sum_{t \in \supp(q_\Scal(T))} q_\Scal(t) \log\left( \frac{q_\Scal(t)}{\qtilde_\Scal(t)} \right). \notag
\end{align}
We also denote by $\Scaltilde \subseteq \Acal$ the support of $\ptilde = \ptilde(A)$.

\begin{proposition} \label{prop:min_lagr_iff_min_lagr_on_S}
  Let $q(T|A) \in C(\Acal, \Tcal)$. Then $q(T|A)$ is a global minimum of $\Lcal_\beta$ if and only If
  \begin{enumerate}
    \item $q_\Scal(T|A)$ is a global minimum of $\Lcal_{\beta,\Scal}$,
    \item For all $t \in \supp(q(T))$ and $a \in \Scaltilde \setminus \Scal$, we have $q(t|a) = 0$.
  \end{enumerate}
  In particular, if $q_\Scal(T|A)$ is a global minimum $\Lcal_{\beta,\Scal}$, we obtain a global minimum of $\Lcal_\beta$ with the extension $q'(T|A) \in C(\Acal, \Tcal)$ of $q_\Scal(T|A)$ defined through
  \begin{align}   \label{eq:loc:def_qprime_TgA}
    q'(T|a) := \begin{cases}
      q_\Scal(T|a) \quad &\text{if $a \in \Scal$}, \\
      \delta_{t_0} \quad &\text{if $a \in \Acal \setminus \Scal$},
    \end{cases}
  \end{align}
  where we chose $t_0 \in \Tcal \setminus \supp(q(T))$.
\end{proposition}
Before proving this result, let us recall that $q(t) = \sum_{a \in \Scal} p(a) q(t|a)$, so that $\supp(q(T)) = \supp(q_\Scal(T))$ can be seen as the ``probabilistic image of $\Scal$ through the channel $q_\Scal(T|A)$'', and does not depend on the values of $q(t|a)$ for $a \in \Scaltilde \setminus \Scal$. Thus the condition $(ii)$ in Proposition~\ref{prop:min_lagr_iff_min_lagr_on_S} means that $q(T|A)$ sends the elements of $\Scal$ and $\Scaltilde \setminus \Scal$ on distinct subsets of bottleneck symbols in $\Tcal$. Moreover, intuitively, the channel $q'(T|A)$ extends $q_\Scal(T|A)$ by sending all the elements $a$ outside $\Scal$ on a ``dummy'' symbol $t_0$ which lies outside the image $\supp(q_\Scal(T))$ of $\Scal$ through $q_\Scal(T|A)$.

\begin{proof}
  We have
  \begin{align}
    D(q(T)||\qtilde(T)) &= \sum_{t \in \supp(q(T))} q(t) \log\left( \frac{q(t)}{\qtilde(t)} \right) \notag \\
    &= \sum_{t \in \supp(q(T)), a \in \Acal} q(t|a) p(a) \log\left( \frac{\sum_{a \in \Acal} q(t|a) p(a)}{\sum_{a \in \Acal} q(t|a) \ptilde(a)} \right) \notag \\
    &= \sum_{t \in \supp(q(T)), a \in \Scal} q(t|a) p(a) \log\left( \frac{\sum_{a \in \Scal} q(t|a) p(a)}{\sum_{a \in \Scal} q(t|a) \ptilde(a) + \sum_{a \in \Scaltilde  \setminus \Scal} q(t|a) \ptilde(a)} \right) \notag \\
    &\leq \sum_{t \in \supp(q(T)), a \in \Scal} q(t|a) p(a) \log\left( \frac{\sum_{a \in \Scal} q(t|a) p(a)}{\sum_{a \in \Scal} q(t|a) \ptilde(a)} \right) \label{eq:loc:dkl_smaller_restricted_dkl} \\
    &= \sum_{t \in \supp(q(T)), a \in \Scal} q'(t|a) p(a) \log\left( \frac{\sum_{a \in \Scal} q'(t|a) p(a)}{\sum_{a \in \Scal} q'(t|a) \ptilde(a)} \right) \notag \\
    &= \sum_{t \in \supp(q(T)), a \in \Acal} q'(t|a) p(a) \log\left( \frac{\sum_{a \in \Acal} q'(t|a) p(a)}{\sum_{a \in \Acal} q'(t|a) \ptilde(a)} \right) \label{eq:loc:dkl_smaller_restricted_dkl_2} \\
    &= \sum_{t \in \Tcal, a \in \Acal} q'(t|a) p(a) \log\left( \frac{\sum_{a \in \Acal} q'(t|a) p(a)}{\sum_{a \in \Acal} q'(t|a) \ptilde(a)} \right) \label{eq:loc:dkl_smaller_restricted_dkl_3}  \\
    &= D(q'(T)||\qtilde'(T)), \notag
  \end{align}
  where we defined with the marginals $q'(T) := \sum_{t \in \Tcal} q'(t|a) p(a)$ and  $\qtilde'(T) := \sum_{t \in \Tcal} q'(t|a) \ptilde(a)$. Note that \eqref{eq:loc:dkl_smaller_restricted_dkl_2} uses $q'(t|a) = 0$ for $a \in \Acal \setminus \Scal,$  $t \in \supp(q(T))$, and \eqref{eq:loc:dkl_smaller_restricted_dkl_3} uses the definition \eqref{eq:loc:def_qprime_TgA} of $q'(T|A)$, which implies
  \begin{align*}
    q'(t_0|a) p(a) = (0 \times p(a)) \, \delta_{a \in \Scal} + (q'(t_0|a) \times 0) \, \delta_{a \in \Acal \setminus \Scal} = 0,
  \end{align*}
  and thus
  \begin{align*}
    &\sum_{a \in \Acal}  q'(t_0|a) p(a) \log \left( \frac{\sum_{a \in \Acal} q'(t_0|a) p(a)}{\sum_{a \in \Acal} q'(t_0|a) \ptilde(a)} \right) = 0.
  \end{align*}

  Moreover, the r.h.s. of \eqref{eq:loc:dkl_smaller_restricted_dkl} coincides with $D(q_\Scal(T)||q_\Scal(T))$.  On the other hand it is straightforward to verify that $I_q(A;T) = I_{q'}(A;T) = I_{q_\Scal}(A;T)$. Thus
  \begin{align} \label{eq:loc:lag_q_larger_lag_qprime}
    \Lcal_\beta(q(T|A)) \geq \Lcal_\beta(q'(T|A)) = \Lcal_{\beta, \Scal}(q_\Scal(T|A)),
  \end{align}
  and equality is achieved in \eqref{eq:loc:lag_q_larger_lag_qprime} if and only if it is achieved in \eqref{eq:loc:dkl_smaller_restricted_dkl}. The latter is equivalent to $\sum_{a \in \Scaltilde  \setminus \Scal} q(t|a) \ptilde(a) = 0$ for all $t \in \supp(q(T))$, i.e., to $q(t|a) = 0$ for all $t \in \supp(q(T))$ and $a \in \Scaltilde \setminus \Scal$, i.e., to point $(ii)$ in Proposition \ref{prop:min_lagr_iff_min_lagr_on_S}.

  Assume now that $q(T|A)$ minimises $\Lcal_\beta$. Then equation \eqref{eq:loc:lag_q_larger_lag_qprime} and its equality case clearly imply point $(ii)$ in Proposition \ref{prop:min_lagr_iff_min_lagr_on_S}. Moreover, if $q_{\Scal,1}(T|A)$ is another channel in $C(\Scal, \Tcal)$, we can extend it to a channel $q_1'(T|A)$ similarly as in \eqref{eq:loc:def_qprime_TgA}, which yields
  \begin{align*}
    \Lcal_{\beta, \Scal}(q_{\Scal,1}(T|A)) = \Lcal_\beta(q_1'(T|A)) \geq \Lcal_\beta(q(T|A)) = \Lcal_\beta(q'(T|A)) = \Lcal_{\beta,\Scal}(q_\Scal(T|A)),
  \end{align*}
  whence point $(i)$ in Proposition \ref{prop:min_lagr_iff_min_lagr_on_S}.

  Conversely, assume that points $(i)$ and $(ii)$ hold. Fix an arbitrary distribution $q_1(T|A) \in C(\Acal, \Tcal)$ and write $q_{\Scal,1}(T|A) \in C(\Scal, \Tcal)$, resp. $q_1'(T|A) \in C(\Acal, \Tcal)$, the restriction of $q_1(T|A)$ to $\Scal$, resp. the corresponding channel defnied similarly as in \eqref{eq:loc:def_qprime_TgA}. Then
  \begin{align*}
    \Lcal_\beta(q_1(T|A)) &\geq \Lcal_\beta(q_1'(T|A)) = \Lcal_{\beta, \Scal}(q_{\Scal,1}(T|A)) \\
    &\geq \Lcal_{\beta,\Scal}(q_\Scal(T|A)) = \Lcal_\beta(q'(T|A)) = \Lcal_\beta(q(T|A)),
  \end{align*}
  where the last equality uses point $(ii)$ and the equality case of \eqref{eq:loc:lag_q_larger_lag_qprime}. Therefore $q(T|A)$ is indeed a global minimum of $\Lcal_\beta$.
\end{proof}

\subsection{Self-consistent equation and Blahut-Arimoto algorithm}
\label{section:apx:blahut_arimoto}

Here we describe a Blahut-Arimoto (BA) iterative algorithm to compute the minimisers of the DIB Lagrangian \eqref{eq:def:sib_lagrangian_apx}. Following Proposition \ref{prop:min_lagr_iff_min_lagr_on_S}, we aim at a minimiser $q_\Scal(T|A)$ of the Lagrangian $\Lcal_{\beta,\Scal}$ restricted to $\Scal := \supp(p(A))$ (see equation \eqref{eq:def:sib_lagrangian_apx_on_S}), which automatically yields solutions for channels defined on the whole alphabet $\Acal$ (see equation \eqref{eq:loc:def_qprime_TgA}). To alleviate notations, in this section we will directly write $q(T|A)$ and $\Lcal_\beta$ instead of $q_\Scal(T|A)$ and $\Lcal_{\beta, \Scal}$. As we will see, our algorithm does not provably converge to a global minimum of the DIB Lagrangian, but it has the same guarantees as the BA algorithm for the classic IB \citep{tishbyInformationBottleneckMethod2000}: namely, the values of the Lagrangian decrease at each step and converge to a fixed value, and the limit of a corresponding convergent sequence $(\kappa_i)_{i \in \mathbb{N}}$ must satisfy equation \eqref{eq:self_consistent_equation}.

\subsubsection{Critical points are characterised by a self-consistent equation}
\label{section:apx:blahut_arimoto_self_consistent}

Taking into account the constraints $\sum_{t \in \Tcal} q(t|a) = 1$ for all $a \in \Scal$, but not the inequality constraints $q(t|a)\geq 0$ for all $a \in \Scal, t \in \Tcal$, we obtain the extended Lagrangian
\begin{align}
    \mathcal{L}_{\beta,\mu}(q(T|A)) :=  I_q(A;T) - \beta D(q(T)||\tilde{q}(T)) + \sum_{a \in \Scal, t \in \Tcal} \mu_{a} q(t|a).
\end{align}
We derive $\mathcal{L}_{\beta,\mu}$ on the open set 
\begin{align*}
  \Qcal_+ := \{ (q(t|a))_{a \in \Scal, t \in \Tcal} \, : \ \forall a \in \Scal, \forall t \in \Tcal, \, q(t|a) > 0 \} = (\mathbb{R}_+)^{|\Scal||\Tcal|}.
\end{align*}
First, $q(t) := \sum_{x'}p(a') q(t|a')$ and $\tilde{q}(t) := \sum_{x'} \ptilde(a') q(t|a')$, so that 
\begin{align*}
    \partial_{q(t|a)} q(t) &= p(a), \\
    \partial_{q(t|a)} \tilde{q}(t) &= \ptilde(a).
\end{align*}
Moreover, note that $q(T)$ and $\qtilde(T)$ are strictly positive for $(q(t|a))_{a,t} \in \Qcal_+$. Thus we can write
\begin{align*}
    \partial_{q(t|a)} I_q(A;T) &= \partial_{q(t|a)} \sum_{a',t'} p(a') q(t'|a') \log\left( \frac{q(t|a')}{q(t)} \right) \\
    &= p(a) \log \left( \frac{q(t|a)}{q(t)} \right) + \sum_{a'} p(a') q(t|a') \frac{q(t)}{q(t|a')} \frac{q(t) \delta_{a' = a} - p(a) q(t|a')}{q(t)^2} \\
    &=  p(a) \log \left( \frac{q(t|a)}{q(t)} \right) + \sum_{a'} p(a') \left( \delta_{a' = a} - \frac{ p(a) q(t|a')}{q(t)} \right) \\
    &=  p(a) \log \left( \frac{q(t|a)}{q(t)} \right) + p(a) - p(a) \frac{q(t)}{q(t)} \\
    &=  p(a) \log \left( \frac{q(t|a)}{q(t)} \right),
\end{align*}
and
\begin{align*}
    \partial_{q(t|a)} D(q(T)||\tilde{q}(T)) &= \partial_{q(t|a)}  \sum_{a',t'} p(a') q(t'|a') \log\left( \frac{q(t)}{\tilde{q}(t)} \right) \\
    &= p(a) \log\left( \frac{q(t)}{\tilde{q}(t)} \right) + \sum_{a'} p(a') q(t|a') \frac{\tilde{q}(t)}{q(t)} \frac{p(a) \tilde{q}(t) - \ptilde(a)q(t)}{\tilde{q}(t)^2} \\
    &= p(a) \log\left( \frac{q(t)}{\tilde{q}(t)} \right) + \left( \sum_{a'} p(a') q(t|a') \right) \left( \frac{p(a)}{q(t)} - \frac{\ptilde(a)}{\tilde{q}(t)} \right) \\
    &= p(a) \log\left( \frac{q(t)}{\tilde{q}(t)} \right) + q(t) \left( \frac{p(a)}{q(t)} - \frac{\ptilde(a)}{\tilde{q}(t)} \right) \\
    &= p(a) \log\left( \frac{q(t)}{\tilde{q}(t)} \right) + p(a) - \frac{q(t)}{\tilde{q}(t)} \ptilde(a).
\end{align*}
Therefore
\begin{align*}
    \partial_{q(t|a)} \mathcal{L}_{\beta,\mu}(q(T|A)) = p(a) \log \left( \frac{q(t|a)}{q(t)} \right) \, - \, \beta \, \left( p(a) \log\left( \frac{q(t)}{\tilde{q}(t)} \right) + p(a) - \frac{q(t)}{\tilde{q}(t)} \ptilde(a) \right) + \mu_{a}.
\end{align*}
Now, recall that here the input set of $q(T|A) = q_\Scal(T|A)$ is $\Scal = \supp(p(A))$. Hence, we can absorb $p(a)$ into the constant $\mu_{a}$, and get that a necessary condition for local minimisers of the DIB Lagrangian $\mathcal{L}_{\beta}$ on $\Qcal_+$ is the existence of constants $(\mu_{a})_a \in \mathbb{R}^{|\Scal|}$ such that
\begin{align*}
    \log \left( \frac{q(t|a)}{q(t)} \right) \, - \, \beta \, \left( \log\left( \frac{q(t)}{\tilde{q}(t)} \right) + 1 - \frac{q(t)\ptilde(a)}{\tilde{q}(t)p(a)} \right) + \mu_{a} = 0
\end{align*}
i.e., such that
\begin{align*}
    & q(t|a) = q(t) \exp\left[ \beta \left( \log\left( \frac{q(t)}{\tilde{q}(t)} \right) + 1 - \frac{q(t)\ptilde(a)}{\tilde{q}(t)p(a)} \right) + \mu_{a}  \right].
\end{align*}
Thus we proved that local minimisers of the DIB Lagrangian $\Lcal_\beta$ over the set of channels $q(T|A) \in C(\Scal, \Tcal)$ with strictly positive entries satisfy the necessary condition
\begin{align}   \label{eq:self_consistent_equation_apx}
    q(t|a) = \frac{1}{Z(a,\beta)} q(t) \exp\left[ - \beta \left( \frac{q(t)\ptilde(a)}{\tilde{q}(t)p(a)} - \log\left( \frac{q(t) \ptilde(a)}{\tilde{q}(t) p(a)} \right) - 1 \right)\right],
\end{align}
where $Z(a,\beta)$ is a positive normaliser. Note that in \eqref{eq:self_consistent_equation_apx}, we added the factor $\frac{\ptilde(a)}{p(a)}$ in the logarithm. This equivalent reformulation is more suited to the implementation of the Blahut-Arimoto algorithm described below. Indeed, in this form, the expression in the exponential is always non-positive (as shown by the study of the function $x \mapsto x - \log(x)-1$), which avoids overflow for large $\beta$.

Note that \emph{a priori}, there might also be local minimisers of $\Lcal_\beta$ on the border of $C(\Scal, \Tcal)$. For the sake of completeness, let us outline an argument showing that this is actually not the case. The computations above show that, deriving $\mathcal{L}_{\beta}(q(T|A))$ as a function on $\Qcal_+$, we get
\begin{align*}
  \partial_{q(t|a)} \mathcal{L}_{\beta}(q(T|A)) = p(a) \left[ \log \left( \frac{q(t|a)}{q(t)} \right) \, - \, \beta \, \left( \log\left( \frac{q(t)}{\tilde{q}(t)} \right) + 1 - \frac{q(t) \ptilde(a)}{\tilde{q}(t)p(a)} \right) \right].
\end{align*}
In particular, for $q(T|A) \in C(\Scal, \Tcal)$ strictly positive but with at least one coordinate approaching $0$, the directional derivative w.r.t this coordinate diverges to $-\infty$. Indeed, $D(p||\ptilde) < +\infty$ implies $\Scal = \supp(p(A)) \subseteq \supp(\ptilde(A))$, while each $q(T|a)$ is a probability, with $p(A)$ and $\ptilde(A)$ fixed; so that there are strictly positive constants $k$ and $K$ such that $k \leq q(t) := \sum_{a} p(a) q(t|a) \leq K$ and $k \leq \qtilde(t) := \sum_{a} \ptilde(a) q(t|a) \leq K$ for all $t \in \supp(q(T))$ and all $q(T|A) \in \Qcal_+$. Thus the term $\log\left( \frac{q(t)}{\tilde{q}(t)} \right) + 1 - \frac{q(t) \ptilde(a)}{\tilde{q}(t)p(a)}$ remains bounded as well. But as $q(t) \geq k$, on the other hand  $\log\left( \frac{q(t|a)}{q(t)} \right)$ diverges to $-\infty$ when $q(t|a)$ goes to $0$.

Using classic arguments, we can then use the divergence to $-\infty$ of the gradient close to the border, along with the continuity of $\Lcal_\beta$ on the whole closed set $C(\Scal, \Tcal)$, to prove that $q(T|A)$ cannot be a local minimum of $\Lcal_\beta$ over $C(\Scal, \Tcal)$ if it has a coordinate $q(t|a)$ equal to $0$, i.e., if it is on the border of $C(\Scal, \Tcal)$.

\subsubsection{Blahut-Arimoto algorithm}
\label{section:apx:blahut_arimoto_algorithm}

Here, we denote by $C_+(\Scal, \Tcal)$ the subset of $C(\Scal, \Tcal)$ made of channels with only positive entries, by $\Delta_{\Tcal, +}$ the open simplex of full-support probabilities on $\Tcal$, and by $\mathbb{R}_+$ the positive real numbers. We define, for $q(T|A) \in C(\Scal, \Tcal)$, a probability $ r(T) \in \Delta_{\Tcal}$ on $\Tcal$, and some $m(T) \in (\mathbb{R}_+)^{|\Tcal|}$,
\begin{align*}
    F(&q(T|A), r(T), m(T)) \\
    & :=  \sum_{a,t} p(a) q(t|a) \log \left( \frac{q(t|a)}{r(t)} \right) - \beta  \sum_{a,t} p(a) q(t|a) \left( \log\left( m(t) \frac{\ptilde(a)}{p(a)} \right) - m(t) \frac{\ptilde(a)}{p(a)} + 1 \right).
\end{align*}
The function $F$ is thus defined on the open and convex set 
\begin{align*}
  \Dom_F := C_+(\Scal, \Tcal) \times \Delta_{\Tcal,+} \times (\mathbb{R}_+)^{|\Tcal|}.
\end{align*}
The next proposition defines the Blahut-Arimoto (BA) algorithm adapted to our problem, and describes its properties.
\begin{proposition} \label{prop:blahut_arimoto}
  The function $F$ is convex in each of its coordinates. Moreover, for $q_i(T|A) \in C_+(\Scal, \Tcal)$, defining
  \begin{align} \label{eq:sib_iterative_eq_full}
    \begin{split}
      r_{i+1}(t) &:= \sum_{a} p(a) q_i(t|a), \\
      m_{i+1}(t) &:= \frac{\sum_{a} p(a) q_i(t|a)}{\sum_{a} \ptilde(a) q_i(t|a)}, \\
      q_{i+1}(t|a) &:= \frac{1}{Z(a,\beta)} r_{i+1}(t) \exp\left[ - \beta \left( m_{i+1}(t) \frac{\ptilde(a)}{p(a)} - \log\left( m_{i+1}(t) \frac{\ptilde(a)}{p(a)} \right) - 1  \right)\right],
    \end{split}
    \end{align}
    where $Z(a,\beta)$ is a positive normaliser, we have:
  \begin{enumerate}
    \item All quantities in \eqref{eq:sib_iterative_eq_full} are well-defined, and $(q_{i+1}(T|A), r_{i+1}(T), m_{i+1}(T)) \in \emph{\Dom}_F$.
    \item $F(q_{i}(T|A), r_{i}(T), m_{i}(T)) = \Lcal_\beta(q_{i}(T|A)) + K$, where the Lagrangian $\Lcal_\beta$ is defined in \eqref{eq:def:sib_lagrangian_apx} and $K$ is a constant that does not depend on $i$.
    \item At each update of $q_i(T|A)$, $r_i(T)$ and $m_i(T)$, the function $F$ is minimised w.r.t. the corresponding coordinate. In particular,
    \begin{align*}
      F(q_{i+1}(T|A), r_{i+1}(T), m_{i+1}(T)) \leq F(q_i(T|A), r_i(T), m_i(T)).
    \end{align*}
  \end{enumerate}
\end{proposition}

Before proving it, let us first draw the consequences of Proposition \ref{prop:blahut_arimoto}. Define some $q_0(T|A) \in C_+(\Scal, \Tcal)$, and the corresponding sequence $(q_i(T|A), r_i(T), m_i(T))_{i \geq 1}$ from \eqref{eq:sib_iterative_eq_full}. From point $(i)$, the sequence is included in $\Dom_F$, and from point $(ii)$, we have, for all $i$,
\begin{align*}
  \Lcal_\beta(q_{i}(T|A)) = F((q_{i}(T|A), r_{i}(T), m_{i}(T))) - K.
\end{align*}
From point $(iii)$, this yields a non-increasing sequence of images $(\Lcal_\beta(q_{i}(T|A)))_i$. As $\Lcal_\beta$ is bounded from below, this implies that this sequence converges. Moreover, as the closure $\overline{C_+(\Scal, \Tcal)} = C(\Scal, \Tcal)$ of $C_+(\Scal, \Tcal)$ is compact, we can, up to extracting a subsequence, assume that $(q_i(T|A))_i$ converges to a point $q_\ast(T|A) \in C(\Scal, \Tcal)$.\footnote{In practice, in numerical implementations, we always observed the convergence of $(q_i(T|A))_i$, without any subsequence extraction.} From the definition of $(q_i(T|A))_i$ through \eqref{eq:sib_iterative_eq_full} and from the continuity of this iterative equation, we obtain that the limit $q_\ast(T|A)$ satisfies the fixed-point equation \eqref{eq:self_consistent_equation_apx}. Hence we proved the claims made the beginning of Appendix \ref{section:apx:blahut_arimoto} about this BA algorithm. Note that even though $F$ is convex in each coordinate, we did not prove that $F$ is convex as a whole. Thus we cannot apply the classic BA arguments \citep{yeungInformationTheoryNetwork2008} to prove that the sequence $(\Lcal_\beta(q_{i}(T|A)))_i$ converges to a global minimum of $\Lcal_\beta$. However, the statements proved here match exactly the corresponding statements proven for the BA algorithm in the classic IB case \citep{tishbyInformationBottleneckMethod2000}.

\begin{proof}[Proof of Proposition \ref{prop:blahut_arimoto}]

  The convexity of $F$ in each coordinate is straightforward. Point $(i)$ comes from the fact that $q(T|A) \in C(\Scal, \Tcal)$, where $\Scal$ is the support of $p(A)$, which contains that of $\ptilde(A)$ (because $D(p||\ptilde) < +\infty)$. Point $(ii)$ is a direct computation. Let us now prove point $(iii)$.
  
  For fixed $(r(T), m(T))$, we know that the function $F(\cdot, r(T), m(T))$ is convex on $C_+(\Scal, \Tcal)$, so that the minimum is achieved at points $q(T|A)$ such that $\nabla_{q(T|A)}F(q(T|A), r(T), m(T) = 0$. A direct computation shows that the latter is equivalent to, for all $a \in \Scal, t \in \Tcal$,
  \begin{align}   \label{eq:iterative_iib_1}
      q(t|a) = \frac{1}{Z(a,\beta)} r(t) \exp\left[ - \beta \left( m(t) \frac{\ptilde(a)}{p(a)} - \log\left( m(t) \frac{\ptilde(a)}{p(a)} \right) - 1  \right)\right],
  \end{align}
  where $Z(a,\beta)$ is a positive normaliser. Moreover, it is standard \citep{yeungInformationTheoryNetwork2008} to prove that, for fixed $(q(T|A), m(T)) \in C_+(\Scal, \Tcal) \times (\mathbb{R}_+)^{|\Tcal|}$, the minimimum of $F$ w.r.t to $r(T)$ is achieved for
  \begin{align}       \label{eq:iterative_iib_2}
      r(T) = q(T):= \sum_{a} p(a) q(T|a)
  \end{align}
  Eventually, for fixed $(q(T|A),r(T)) \in C_+(\Scal, \Tcal) \times \Delta_{\Tcal,+}$ the minimum of $F$ w.r.t. $m(T)$ is, again by convexity, achieved if and only if the corresponding gradient vanishes. But we have
  \begin{align*}
      \partial_{m(t)} F(q(T|A), r(T), m(T)) &= \sum_{a} p(a) q(t|a) \left( \frac{1}{m(t)} - \frac{\ptilde(a)}{p(a)} \right) \\ 
      &= \frac{q(t)}{m(t)} - \qtilde(t) ,
  \end{align*}
  so that the gradient w.r.t $m(T)$ cancels if and only if for all $t \in \mathcal{T}$,
  \begin{align}   \label{eq:iterative_iib_3}
      m(t) = \frac{q(t)}{\qtilde(t)} = \frac{\sum_{a} p(a) q(t|a)}{\sum_{a} \ptilde(a) q(t|a)}.
  \end{align}
  This proves point $(iii)$.
\end{proof}

\subsection{Details on effective cardinality}
\label{section:apx:charac_effcard_with_frac}

\citep{zaslavskyDeterministicAnnealingEvolution2019} defines a concept of effective cardinality for the Lagrangian formulation of the classic IB. Here, we adapt this concept to the DIB framework in its primal formulation, i.e., problem \eqref{eq:def:sib}, and in a way which also encompasses the case $\supp(p(A)) \subsetneq \Acal$. For $\kappa = q(T|A) \in C(\Acal, \Tcal)$, consider the ``probabilistic image of $\Acal$ through $\kappa$'', i.e.,
\begin{align*}
  \kappa \cdot \Acal := \{ t \in \Tcal \, : \ \exists a \in \Acal \, : \, q(t|a) > 0 \}.
\end{align*}
Note that this definition depends only on $\kappa$ and not on $p(A)$. We then define the cardinality of  $\kappa$ as $K(\kappa) := |\kappa \cdot \Acal|$. However, for $\kappa \in \iname(\lambda)$, the number $K(\kappa)$ does not necessarily carry any meaningful information about the DIB problem itself: e.g., it can be easily verified that composing any $\kappa \in \iname(\lambda)$ with a congruent channel $\gamma$ (which can arbitrarily increase the cardinality $K(\kappa)$) still yields a solution $\gamma \circ \kappa \in \iname(\lambda)$. This motivates the definition of the minimum number of symbols $t$ necessary to describe the output of a bottleneck encoder $\kappa$. Formally:
\begin{definition}
  The \emph{effective cardinality} of a $\inameop$ solution $\kappa \in \inameop(\lambda)$ is
  \begin{align*}
    k(\kappa) := \min_{\gamma \in C(\Tcal) : \ \gamma \circ \kappa \in \inameop(\lambda)} \, K(\gamma \circ \kappa),
  \end{align*}
  i.e., it is the minimum bottleneck cardinality obtained from a post-processing of $\kappa$ that still produces a $\inameop$ solution for the same parameter $\lambda$.
\end{definition}
Let us fix an arbitrary $\kappa = q(T|A) \in \iname(\lambda)$, write $q(A,T) :=  p(A) q(T|A)$, and assume first that $p(A)$ is full support. It can be shown, using the log-sum inequality, that $k(\kappa)$ is the cardinality of the partition $\bar{\Tcal}$ of $\supp(q(T))$ defined by the equivalence relation $t \sim t' \Leftrightarrow q(A|t) = q(A|t')$.

For the non full support case, denote by $\ptilde(A)$ the unique distribution satisfying $D(p||\ptilde) = D(p||\Ecal)$ \citep{ayInformationGeometry2017}, and note that $D(p||\Ecal) < \infty$ implies $\Scal \subseteq \Scaltilde$, where $\Scal := \supp(p(A))$ and $\Scaltilde := \supp(\ptilde(A))$. It can be easily verified that the value of $q(T|a)$ for $a \in \Scaltilde^c$ affects neither the target nor the constraint function of the DIB problem \eqref{eq:def:sib}. However, a direct consequence of Proposition~\ref{prop:min_lagr_iff_min_lagr_on_S} is that if $\Scaltilde \setminus \Scal \neq \emptyset$, the image $\kappa \cdot \Acal$ of $\Acal$ through a solution $\kappa \in \iname(\lambda)$ must contain at least one symbol $t_0 \notin \kappa \cdot \Scal$, on which to send the elements of $\Scaltilde \setminus \Scal$. Here we denoted by $\kappa \cdot \Scal$ the ``probabilistic image of $\Scal$ through the channel $\kappa = q(T|A)$'', i.e.,
\begin{align*}
  \kappa \cdot \Scal := \{ t \in \Tcal \, : \ \exists a \in \Scal \, : \, q(t|a) > 0 \}.
\end{align*}
Note that $\kappa \cdot \Scal = \supp(q(T))$, for $q(T) := \sum_{a \in \Scal} p(a) q(T|a)$.

It can be easily verified that the previous paragraph implies that in the non full support case, the effective cardinality becomes $|\bar{\Tcal}| + 1$. Note that this is the situation we encounter in our numerical experiments (Section \ref{section:numerical_experiments_equivariance}).

We use the above to numerically compute the effective cardinality. Note that the choice of the threshold for rounding $|q(t|a) - q(t|a')|$ to $0$ is here important. We choose $10^{-3}$.

\subsection{Computable form of $D^p(\kappa||C_G$)}
\label{section:apx:computation_channel_divergence}

Here we provide more details on the divergence introduced in Section \ref{section:sib_tools}, and prove that it can be computed directly as a divergence between two channels. Let $\Scal := \supp(p(A))$. For two channels $\kappa, \nu$ in either $C(\Acal, \Tcal)$ or $C(\Scal, \Tcal)$, we define their Kullback-Leibler divergence $D^p(\kappa||\nu)$ with respect to $p = p(A)$, as \citep{ayInformationGeometry2017}
\begin{align*}
    D^p(\kappa||\nu):= \sum_{a \in \Scal} p(a) D(\kappa(T|a)||\nu(T|a)).
\end{align*}
We also define, for a group $G$ acting on $\Acal$, the set of input-symmetric channels w.r.t. $G$, i.e.,
\begin{align*}
  C_{G} &:= \{ \nu \in C(\Acal, \Tcal) : \quad \forall \Phi \in G, \ \ \nu \circ \Phi = \nu \  \},
\end{align*}
and the corresponding divergence of some $\kappa \in C(\Acal, \Tcal)$ from $C_G$ with respect to $p$ as \citep{ayInformationGeometryComplexity2015}
\begin{align*}
  D^p(\kappa || C_G) := \min_{\nu \in C_G} \, D^p(\kappa || \nu).
\end{align*}
For all purposes relevant to this article's scope, we have $D^p(\kappa || C_G)$ if and only if $\kappa \circ \Phi = \kappa$ for all $\Phi \in G$. More precisely:

Assume first that $p(A)$ is full support. Then, from the continuity of the KL divergence and the fact that $C_G$ is a closed subset of $C(\Acal,\Tcal)$, we have $D(\kappa || C_G) = 0$ if and only if $\kappa \in C_G$, i.e., $\kappa \circ \Phi = \kappa$ for all $\Phi \in G$.

Let us now drop the full support assumption on $p(A)$, but assume instead that $(i)$ the group $G$ leaves $\Scal$ invariant, and $(ii)$ the channel $\kappa = q(T|A)$ is as $q'(T|A)$ in equation \eqref{eq:loc:def_qprime_TgA}, i.e., it sends $\Scal^c$ on a single symbol outside the image of $\Scal$ through $\kappa$. From point $(i)$, the action of $G$ on $\Acal$ induces an action on $\Scal$, and a corresponding set $C_{G_\Scal}$. Denote by $\kappa_\Scal$ the restriction of a channel $\kappa \in C(\Acal, \Tcal)$ to $\Scal$. Using both points $(i)$ and $(ii)$, it can be easily verified that for all $\kappa \in C(\Acal, \Tcal)$, we have $\kappa \in C_G$ if and only if $\kappa_\Scal \in C_{G_\Scal}$, and that  $D^p(\kappa || C_{G}) = D^p(\kappa_\Scal || C_{G_\Scal})$. From that we can conclude, using the full support case described above, that here we also have $D^p(\kappa || C_G)$ if and only if $\kappa \circ \Phi = \kappa$ for all $\Phi \in G$.

Note that points $(i)$ and $(ii)$ are satisfied in our numerical experiments in Section \ref{section:numerical_experiments_equivariance}, and that they are also automatically satisfied if $p(A)$ is full support.

Let us now provide a form of $D^p(\kappa || C_G)$ which is easier to compute.

\begin{proposition} \label{prop:divergence_from_input_symmetric_projection}
  Fix $p(A) \in \Delta_{\Acal}$, a finite group $G$ acting on $\Acal$ and leaving $\Scal$ invariant, and $\kappa = q(T|A) \in C(\Acal, \Tcal)$. Then
  \begin{align*}
      D^p(\kappa||C_G) =   D^p(\kappa||\kappa_G)
  \end{align*} 
  where $\kappa_G = q_G(T|A) \in C(\Scal, \Tcal)$ is defined through, for $a \in \Scal$ and $t \in \Tcal$,
  \begin{align*}
    {q_G}(t|a) := q(t|[a]) := \frac{\sum_{a' \in [a]} p(a') q(t|a')}{p([a])},
  \end{align*}
  with $[a]$ the orbit of $a$ under $G$.
\end{proposition}
Intuitively, $\kappa_G$ is the average of the channel $\kappa$ over the group $G$ acting on its input, computed using the distribution $p$ on the input. 

\begin{proof}
  It is enough to prove that for all $\nu \in C_G$,
  \begin{align*}
      D^p(\kappa||{\kappa_G}) \leq D^p(\kappa||\nu).
  \end{align*}
  For $a \in \Scal$, we have $[a] \subseteq \Scal$ (because $G$ leaves $\Scal$ invariant), and $q_G(T|a')$ is well-defined and constant for $a' \in [a]$. Moreover, for $\nu = r(T|A) \in C_G$, it is straightforward to verify that $r(T|a')$ is also constant for $a' \in [a]$, so that
  \begin{align*}
    \sum_{a' \in [a]} r(t|a') p(a') = r(t|a) p([a]).
  \end{align*}
  Thus, for $\nu = r(T|A) \in C_G$, and $a_1, \dots, a_n$ a system of representatives of all the orbits included in $\Scal$,
  \begin{align*}
      D^p(\kappa||\nu) - D^p(\kappa||{\kappa_G}) &= \sum_{a \in \Scal, t \in \Tcal} p(a) q(t|a) \log \left( \frac{q_G(t|a)}{r(t|a)} \right) \\
      &= \sum_{i=1}^n \sum_{t} \log \left( \frac{q_G(t|a_i)}{r(t|a_i)} \right) \sum_{a \in [a_i]} p(a) q(t|a)  \\
      &= \sum_{i=1}^n \sum_{t} \log \left( \frac{\sum_{a \in [a_i]} q(t|a) p(a)}{\sum_{a \in [a_i]} r(t|a) p(a)} \right) \sum_{a \in [a_i]} p(a) q(t|a) \\
      &=  D(q_1||q_2) \geq 0,
  \end{align*}
  where $q_1$ and $q_2$ are distributions defined on $\nicefrac{\Scal}{G} \times \Tcal$, through
  \begin{align*}
      q_1([a_i],t) = \sum_{a \in [a_i]} p(a) q(t|a), \\
      q_2([a_i],t) = \sum_{a \in [a_i]} p(a) t(t|a).
  \end{align*}
\end{proof}

\end{document}